\documentclass{jpp}
\usepackage{graphicx}
\usepackage{subcaption}
\usepackage{booktabs}
\usepackage{siunitx}
\usepackage{rotating}
\usepackage{amssymb}
\usepackage[utf8]{inputenc}
\usepackage[T1]{fontenc}
\usepackage{amsmath}
\usepackage{url}

\usepackage{soul}
\usepackage[dvipsnames]{xcolor} 

\shorttitle{Landau damping closure via Neural operator}
\shortauthor{S. Burles, E. Camporeale and O. Pezzi}

\title{A Neural Operator Closure for Landau Damping in Electrostatic Plasma}

\author{Samuel Burles\aff{1}
  \corresp{\email{s.burles@qmul.ac.uk}},
  E. Camporeale\aff{1,2}
 \and O. Pezzi\aff{3}}

\affiliation{\aff{1}School of Physical and Chemical Sciences, Queen Mary University of London, London, E1 4NS, UK
\aff{2}Space Weather TREC, University of Colorado, Boulder, CO 80303, USA
\aff{3}Institute for Plasma Science and Technology, Consiglio Nazionale delle Ricerche
Via G. Amendola 122/D, 70126 Bari, Italy}

\begin{document}

\maketitle
\begin{abstract}
We present a data-driven plasma fluid closure for both linear and nonlinear electrostatic Landau damping in one dimension. A Fourier Neural Operator (FNO) is trained \textit{online} within a differentiable fluid solver, with the loss computed on trajectories produced by the closed fluid simulation rather than on individual kinetic snapshots. The closure is non-Markovian, acting on a trailing window of the resolved moment history so as to represent the memory of the unresolved dynamics. We demonstrate that a single FNO trained in this way reproduces both linear and nonlinear Landau damping, generalises to initial perturbation amplitudes outside the training set, and remains numerically stable when deployed in independent fluid simulations. In the nonlinear regime the learned heat flux reproduces the resolved-moment dynamics without matching the kinetic heat flux pointwise, behaving as an effective closure that compensates for the truncated higher moments, though the learned specific flux is expected to depend on the numerical scheme and training data. A sensitivity analysis of the trained model shows that it computes a genuine moment-to-flux relation whose reliance on the memory window is physically structured.
\end{abstract}

\section{Introduction}

Collisionless plasma dynamics is multi-scale. Kinetic effects influence large-scale structure, and accurate global simulation of a large-scale system therefore requires resolving kinetic interactions across all scales \citep{lapenta_we_2022, lautenbach_multiphysics_2018}. A quote from Krall and Trivelpiece summarises this well: ``The microscopic quantities are more difficult to measure directly, but frequently they play a dominant role in determining the macroscopic properties of plasma'' \citep{krall_principles_1973}.

Historically, efforts to model collisionless plasma have been split between two approaches: kinetic models and fluid models. Kinetic simulations resolve particle-level interactions, but become computationally prohibitive at large scales due to the high dimensionality of the phase space. Fluid models of plasma, obtained through reductions of the kinetic description, are much more computationally viable by virtue of their low dimensionality, but are ill-suited to resolving kinetic effects because of the simplifying approximations required to construct them.

Indeed, fluid models can be obtained by taking velocity moments of the Vlasov equation. The evolution of each order of velocity moment depends on the next-higher moment, producing an infinite hierarchy of equations that must be truncated at some chosen order. Truncation requires a closure relation that expresses the chosen moment in terms of lower-order moments only. Closures are therefore approximations, and the quality of this approximation determines how much of the underlying kinetic physics the resulting fluid model can recover.

The development of collisionless plasma closures began with the seminal closure of Chew, Goldberger and Low (CGL closure) \citep{chew_boltzmann_1956}, which introduced pressure anisotropy to collisionless plasma fluid models and gave rise to the adiabatic plasma fluid models \citep{le_hybrid_2016, le_equations_2009, ohia_demonstration_2012}. Substantial subsequent effort has focused on the phenomenon of Landau damping, beginning with the Hammett-Perkins closure \citep{hammett_fluid_1990}, which describes the linear phase of Landau damping through a closure formulated in Fourier space and derived from linear kinetic theory. The wider family of fluid models incorporating Landau damping, known as Landau fluids, has developed from this initial work by Hammett and Perkins \citep{snyder_landau_1997, ng_improved_2020, wang_comparison_2015, goswami_landau_2005, passot_fluid_2006, sulem_landau_2015}, including systematic higher-order closures constructed from Padé approximants of the plasma dispersion function \citep{hunana_introductory_2019, hunanaNewClosuresMore2018}, which improve the fidelity of the linear kinetic response across a wider range of phase velocities. Closures derived from linear theory, such as Hammett–Perkins and its higher-order Padé descendants, are by construction limited to the regimes in which that theory applies, motivating data-driven approaches that can extend beyond them.

Recent advances in parallel computing and GPU hardware have driven rapid progress in machine learning, resulting in new techniques applied across the physical sciences. In neutral fluid dynamics, machine learning approaches have been adopted with particular enthusiasm for turbulent simulation and the closure problem more broadly \citep{gupta_generalized_2023, li_synthetic_2024, brunton_machine_2020, sanderse_scientific_2024}.

In plasma physics, early efforts to apply machine learning to the closure problem focused on reproducing known analytic closures, such as the aforementioned Hammett-Perkins closure, with neural-network surrogate models, establishing that closure relations can in principle be represented by learned models \citep{ma_machine_2020, maulik_neural_2020}. Subsequent work moved beyond analytic targets and trained closures directly on data from fully-kinetic simulations \citep{huang_machine-learning_2025, qin_data-driven_2023, liu_data-driven_2022, joglekar_machine_2023, mcgrae-menge_embedding_2026} to capture physics that lies outside the range of validity of any single analytic closure theory. More recent studies have extended this approach to higher-dimensional settings and to richer phenomena, including magnetic reconnection and plasma turbulence \citep{laperre_identification_2022, miloshevich_electron_2025}. We present a much more detailed review of the field in \citet{burles_machine_2026}.

Two recent studies are of particular relevance to the present work. \citet{wei_data-driven_2023} presented the first attempt at using the Fourier Neural Operator (FNO) architecture to reproduce the heat flux gradient from a Vlasov-Amp\'{e}re simulation, showing that an FNO more accurately reproduces the kinetic heat flux when compared to a more typical multilayer perceptron (MLP) architecture. \citet{huang_machine-learning_2025} later demonstrated that an FNO can serve as a closure for one-dimensional Landau damping, reproducing both the linear and nonlinear regimes when deployed within a fluid solver. Their closure is trained \textit{offline} (fitted as a regression problem against moments pre-computed from the kinetic simulation data) and subsequently tested \textit{online} (the trained model is coupled with a fluid solver and performance is judged at inference). The closure from \citet{huang_machine-learning_2025} demonstrated the viability of this method for including nonlinear Landau damping in a fluid system; however, their results were limited to only the initial damping and growth phase of the nonlinear simulation.

This offline construction is an instance of what the neutral-fluid and climate closure-modelling communities call \textit{a priori} training, in which the model is fitted to reproduce the true closure term on precomputed data, as opposed to \textit{a posteriori} training, in which the model is embedded in a differentiable solver and optimised against the resulting trajectories \citep{frezat_posteriori_2022, macart_embedded_2021, shankar_differentiable_2024, sanderse_scientific_2024}. A recurring finding of that work is that a-priori-optimal closures can be unstable once coupled to the solver, whereas a-posteriori training tends to yield closures that are more stable and accurate in deployment \citep{frezat_posteriori_2022}; the online training we employ here is the plasma-closure counterpart of this a-posteriori strategy. This leaves two questions unresolved. First, low pointwise error on kinetic snapshots does not guarantee numerical stability once the closure is embedded in a fluid solver, where small per-step errors can compound over thousands of timesteps and drive the simulation away from the physical solution. Concrete demonstrations of learned closures in genuine \textit{online} tests remain scarce \citep{huang_machine-learning_2025, joglekar_machine_2023, shukla_learned_2022}.

Second, for a closure to be useful in large-scale fluid modelling it must generalise across initial conditions, ideally capturing the linear and nonlinear regimes of a given process within a single fluid model. Existing studies typically demonstrate accuracy at one or a few initial conditions rather than across a span of initial conditions. The question of how reliably a learned closure transfers to unseen initial states has not been systematically addressed.

In this work we address both of these gaps. We embed an FNO closure directly within a differentiable fluid solver and train it online, computing the loss on trajectories produced by the closed fluid simulations rather than on individual kinetic snapshots. By training the closure in the same setting in which it is subsequently deployed, online training yields a model that is, by construction, suited to stable use in independent fluid simulation. We then demonstrate that a single FNO closure trained in this manner reproduces both linear and nonlinear electrostatic Landau damping in one dimension, and that it generalises to initial perturbation amplitudes outside of the training distribution. This represents a concrete step towards a broadly applicable machine-learned plasma closure, capable of acting as a component within larger fluid simulations across the range of initial conditions such simulations encounter.

A distinguishing feature of our closure is that it is \emph{non-Markovian}: rather than mapping the instantaneous resolved moments to the heat flux, it acts on a trailing window of the recent moment history. This choice is motivated by the observation that the exact closure obtained by projecting out the unresolved dynamics carries memory of the resolved fields, as made precise by the Mori-Zwanzig formalism, in which the unresolved contribution appears as a memory integral over the history of the resolved variables \citep{gouasmi_priori_2017, parish_nonmarkovian_2017}. Finite-memory neural closures of this kind have been explored for reduced-order models of neutral fluids \citep{wang_recurrent_2020}; here the memory encodes the phase-mixing history that underlies Landau damping. Having trained the closure, we further probe what it has learned through a sensitivity analysis of its inputs, providing evidence that it computes a genuine moment-to-flux relation rather than extrapolating its own output, and that its reliance on the memory window is physically structured.

We choose Landau damping of a Langmuir wave as a test case for this method as it is well established in the existing machine-learned closure literature, and provides a simple case of kinetic dynamics with well-known behaviour in both the linear and nonlinear regimes. The two regimes are separated by the trapping (bounce) time $\tau_{\mathrm{tr}} = (m / e E_0 k)^{1/2}$, after which the linear theory of Landau damping breaks down; here $k$ is the wavenumber and $E_0$ is the amplitude of the electric field \citep{oneil_collisionless_1965}. For a fixed wavenumber this timescale is set by the perturbation amplitude, so scanning the initial amplitude, as we do here, moves the system continuously from the linear ($t \ll \tau_{\mathrm{tr}}$) to the trapping-dominated nonlinear ($t \gtrsim \tau_{\mathrm{tr}}$) regime.

The remainder of the paper is organised as follows. Section~\ref{sec:background} reviews the general plasma closure problem and sets out the one-dimensional electrostatic fluid models, framing the closure problem in the form used here. Section~\ref{sec:methods} describes the FNO architecture and the online training procedure within the differentiable fluid solver. Section~\ref{sec:results} presents the results for an FNO trained on single initial perturbation amplitudes representing a linear case and a nonlinear case of Landau damping, the generalisation of a single FNO trained across multiple initial perturbation amplitudes, a comparison between the learned effective heat flux and the kinetic heat flux, and a sensitivity analysis probing what the closure has learned. Section~\ref{sec:discussion} discusses the implications of our findings and outlines directions for future work.
 
\section{Plasma Closure Background}
\label{sec:background}

Following the procedure set out in \cite{swanson_plasma_2003}, we write the collisionless Vlasov equation as
\begin{equation}
    \frac{\partial f_\alpha}{\partial t} + \boldsymbol{v} \cdot \bnabla f_\alpha + \frac{q_\alpha}{m_\alpha} \left( \boldsymbol{E} + \boldsymbol{v} \times \boldsymbol{B} \right) \cdot \bnabla_{\boldsymbol{v}} f_{\alpha} = 0.
\end{equation}

Here $f_\alpha(\boldsymbol{x}, \boldsymbol{v}, t)$ is the phase-space distribution function of species $\alpha$, $\boldsymbol{v}$ is the velocity, $q_\alpha$ and $m_\alpha$ are the charge and mass of the species, $\boldsymbol{E}$ and $\boldsymbol{B}$ are the electric and magnetic fields, and $\bnabla$ and $\bnabla_{\boldsymbol{v}}$ denote gradients in configuration and velocity space. The subscripts $\alpha$ denote the particle species, which we will henceforth drop as we consider a one-fluid model of electrons with a static ion background distribution. We introduce a scalar function of velocity, $\phi_{j}(\boldsymbol{v}) = \boldsymbol{v}^{\,j}$, of moment order $j$, and define the moment-taking process as averaging over the velocity:
\begin{equation}
    \langle\phi_{j}(\boldsymbol{v})\rangle \; = \; \int \phi_{j} f \, \mathrm{d}\boldsymbol{v}.
    \label{average_def}
\end{equation}
We then proceed to take a general velocity moment of the Vlasov equation in order to obtain the evolution equation for a general-order moment.
\begin{equation}
    \frac{\partial}{\partial t} \langle\phi_{j}\rangle + \bnabla \cdot \langle \phi_{j} \boldsymbol{v} \rangle - \langle \boldsymbol{A} \cdot \bnabla_{\boldsymbol{v}} \phi_{j} \rangle = 0,
\end{equation}
where $\boldsymbol{A} = \frac{q_{\mathrm{e}}}{m}(\boldsymbol{E} + \boldsymbol{v} \times \boldsymbol{B})$, $q_{\mathrm{e}}=-e$ is the electron charge (with $e$ the elementary charge) and $m$ is the electron mass. It is clear from the second term in this equation that the evolution of each moment is dependent on the next highest-order moment, resulting in an infinite BBGKY hierarchy of moment evolution equations that must be truncated in order to obtain a fluid model. A closure relation is an equation that approximates the highest-order moment retained in a fluid model using only lower-order moments.

Henceforth, we restrict ourselves to the case of a one-dimensional, electrostatic electron plasma with a static ion background, as this will be our setup for investigating a closure relation for Landau damping. Fluid models of plasma are obtained by taking velocity moments of the Vlasov equation. The fluid variables are defined as the velocity moments of the one-dimensional electron distribution function $f(x,v,t)$.
\begin{align}
    n(x,t) &= \int f(x,v,t) \, \mathrm{d}v, \\
    u(x,t) &= \frac{1}{n} \int v f(x,v,t) \, \mathrm{d}v, \\
    p(x,t) &= m \int (v - u)^2 f(x,v,t) \, \mathrm{d}v, \\
    q(x,t) &= m \int (v - u)^3 f(x,v,t) \, \mathrm{d}v.
\end{align}
Here $n$ is the number density, $u$ the bulk (mean) velocity, $p$ the scalar pressure and $q$ the heat flux, and $m$ is the electron mass as above.

Applying the moment-taking procedure, up to second order, to the Vlasov equation in our one-dimensional electrostatic setting yields the fluid evolution equations
\begin{align}
    \frac{\partial n}{\partial t} &= - \frac{\partial}{\partial x} (nu), \label{eq:continuity}\\
    \frac{\partial u}{ \partial t} &= - u \frac{\partial u}{\partial x} - \frac{1}{mn} \frac{\partial p}{\partial x} + \frac{q_{\mathrm{e}}}{m} E, \label{eq:momentum}\\
    \frac{\partial p}{\partial t} &= - u \frac{\partial p}{\partial x} - 3 p \frac{\partial u}{\partial x} - \frac{\partial q}{\partial x}. \label{eq:pressure}
\end{align}
Equation~(\ref{eq:continuity}) is the continuity equation, Equation~(\ref{eq:momentum}) is the momentum equation closed by the electric field, and Equation~(\ref{eq:pressure}) is the pressure evolution equation. The system is closed electrostatically through Poisson's equation, which relates the electric field to the departure of the electron density ($n(x,t)$) from the neutralising ion background ($n_0$),
\begin{equation}
    \frac{\partial E}{\partial x} = \frac{q_{\mathrm{e}}}{\epsilon_0}(n_0 - n(x,t)).
    \label{eq:poisson}
\end{equation}
The remaining closure problem is to express the third-order moment $q$, or equivalently its gradient $\partial q / \partial x$, as a function of the lower-order fluid quantities $n$, $u$, $p$ and the electric field $E$.

The best-known analytic solution to this closure problem is the Hammett-Perkins closure \citep{hammett_fluid_1990}, obtained by matching the linear fluid response to the kinetic plasma dispersion function. In Fourier space it expresses the heat flux as a spatially non-local operator on the temperature perturbation,
\begin{equation}
    \tilde{q}_k = -n_0 \sqrt{\tfrac{8}{\upi}} v_{\mathrm{th}} \frac{\mathrm{i}k}{|k|} \tilde{T}_k,
    \label{eq:hp}
\end{equation}
with $v_{\mathrm{th}} = \sqrt{T_0/m}$ and $\tilde{T}_k$ the Fourier transform of the temperature perturbation $\tilde{T} = \tilde{p} - T_0 \tilde{n}$. The factor $\mathrm{i}k/|k| = \mathrm{i} \, \mathrm{sgn}(k)$ is a Hilbert transform in configuration space, making the closure non-local in space but instantaneous in time and linear in the moments. Higher-order Landau-fluid closures generalise (\ref{eq:hp}) by approximating the plasma dispersion function with higher-order Pad\'{e} approximants \citep{hunanaNewClosuresMore2018, hunanaIntroductoryGuideFluid2019}, improving the linear response over a broader range of phase velocities while remaining, like Hammett-Perkins, closures derived from linear kinetic theory. We take the Hammett-Perkins closure as an analytic benchmark for the learned closure (Sections~\ref{sec:linear}, \ref{sec:nonlinear}).

\section{Methods}
\label{sec:methods}

Our approach proceeds in three stages: generation of kinetic simulation data, online training of the FNO within a differentiable fluid solver, and online testing of the trained model in an independent fluid simulation.

Kinetic simulations are run using a Vlasov-Poisson solver to produce noise-free distribution function outputs, from which we compute velocity moments. During online training, the FNO output replaces the heat-flux gradient in Equation~(\ref{eq:pressure}). Equations (\ref{eq:continuity}-\ref{eq:pressure}) are solved via a differentiable fluid solver and a loss function is defined as the difference between fluid and kinetic moments computed over a simulation window (so-called rollout window). This procedure trains the FNO to produce a heat-flux gradient that yields fluid trajectories matching the kinetic ones, without ever explicitly fitting the FNO to the kinetic heat flux itself. Finally, the trained FNO is tested online by deploying it within a fluid solver initialised from the same initial conditions as the corresponding kinetic simulation, and the resulting trajectories are compared with the kinetic ground truth over the full simulation duration.

\subsection{Kinetic Simulation Data}

We employ an Eulerian Vlasov-Poisson algorithm to generate the kinetic simulation data \citep{2013PhPl...20i2111P, pezzi_collisional_2016,2023PhPl...30i2304C}. Eulerian Vlasov solvers are particularly suitable for our purposes and , hence, preferred over Particle-In-Cell approaches, for our purposes because they discretise the plasma distribution function directly on a phase-space grid. This ensures an extremely clean description of the distribution function --compared to the inherent statistical noise of Particle-In-Cell methods--, and hence of the velocity moments computed from these outputs. The algorithm numerically integrates Vlasov equation for electrons --assuming that ions are a static background-- coupled to the Poisson equation. Vlasov equation is solved implementing a splitting scheme \citep{CHENG1976330,FILBET20021} that splits the multi-dimensional advection in phase space into several separate one-dimension advections in physical and velocity space. Each one-dimensional advection is then performed through the fourth-order volume-difference Van Leer scheme \citep{MANGENEY2002495}. It assumes periodic boundary conditions in physical space, while in velocity space $f(|v|>v_{\mathrm{max}})=0$, where $v_{\mathrm{max}}$ is a large multiple of the electron thermal speed to ensure a neat description of the distribution function. The code also implements collisional effects by introducing opportune Fokker-Planck operators at the right-hand side of the Vlasov equation. However, for this study we focus on collisionless simulations. 

The simulation domain is one-dimensional with periodic boundary conditions, with system length $L = 2\upi / k$ where $k = 0.35$ is the wavenumber of the initial perturbation, such that the simulation length is normalised to a single period of the initial perturbation. The kinetic simulation uses 256 grid points in configuration space and 401 grid points in velocity space, spanning $v \in [-6,6]$, normalised by the electron thermal velocity. Configuration-space data is downsampled to 128 grid points before being supplied to the fluid solver during training. In both the kinetic and fluid simulations, the simulation time is normalised by the plasma frequency, $\omega_{\mathrm{p}}$. The simulation is initialised with a static ion background distribution and a perturbed electron distribution; the perturbation is sinusoidal in space and is varied in amplitude across simulation runs in order to probe both the linear and nonlinear regimes of Landau damping. The initial perturbation amplitude $A$ enters as a perturbation to the electron density,
\begin{equation}
    n_0(x) = 1 + A \cos (kx).
\end{equation}
From each kinetic snapshot, we compute the velocity moments up to and including the heat flux using Simpson's rule applied to the velocity-space integrals defined in Section~\ref{sec:background}. These moment snapshots serve both as the input initial conditions, evolved during training in the fluid rollouts, and as the ground-truth targets against which the closed fluid trajectories are compared. The parameters of the kinetic simulations are collected in Table~\ref{tab:kinetic_params}.

\begin{table}
  \centering
  \begin{tabular}{lc}
    \toprule
    Parameter & Value \\
    \midrule
    Spatial grid points, $N_x$          & 256 \\
    Velocity grid points, $N_v$         & 401 \\
    Velocity domain                     & $v \in [-6, 6]\,v_{\mathrm{th}}$ \\
    Perturbation wavenumber, $k$        & 0.35 \\
    Domain length, $L$                  & $2\upi/k$ \\
    Boundary conditions                 & periodic \\
    Ion background                      & static, uniform \\
    Initial density                     & $n_0(x) = 1 + A\cos(kx)$ \\
    Timestep, $\Delta t_{\mathrm{kin}}$ & $10^{-3}\,\omega_{\mathrm{p}}^{-1}$ \\
    Output cadence                      & $0.1\,\omega_{\mathrm{p}}^{-1}$ (every 100 steps) \\
    Run duration (linear)               & $150\,\omega_{\mathrm{p}}^{-1}$ \\
    Run duration (nonlinear)            & $500\,\omega_{\mathrm{p}}^{-1}$ \\
    \bottomrule
  \end{tabular}
  \caption{Parameters of the kinetic (Vlasov) simulations used to generate the training and ground-truth data. Velocities are normalised by the electron thermal speed $v_{\mathrm{th}}$, lengths by the Debye length, and time by the inverse plasma frequency $\omega_{\mathrm{p}}^{-1}$. Configuration-space fields are downsampled from $N_x=256$ to the $N_G=128$ fluid grid before being supplied to the closure.}
  \label{tab:kinetic_params}
\end{table}

\subsection{Fluid Model}
\label{sec:fluid_model}

In our strategy, following prior works by \citet{huang_machine-learning_2025} and \citet{wei_data-driven_2023}, the FNO is incorporated directly into the fluid equations, replacing the heat-flux-gradient term in the pressure equation~(\ref{eq:pressure}). Rather than treating the closure as an instantaneous algebraic map from the current lower-order moments to the heat-flux gradient, we allow the closure to depend on a short \emph{history} of the resolved fields. The closure takes as input the trailing window of the $K$ most recent states, together with an amplitude-conditioning channel $\log A_{\mathrm{scale}}$ (defined in Section~\ref{sec:fno}), and returns the heat-flux gradient at the next step. The pressure equation becomes
\begin{equation}
  \frac{\partial p}{\partial t}
  = -\,u\frac{\partial p}{\partial x}
    -\,3p\frac{\partial u}{\partial x}
    -\,\mathrm{FNO}\!\left(\mathbf{M};\,\log A_{\mathrm{scale}}\right).
  \label{eq:closure}
\end{equation}
where $\mathbf{M}$ is defined below. The use of a finite memory window is physically motivated. The exact closure obtained by projecting out the unresolved (higher-moment, fine velocity-space) dynamics is in general non-Markovian: the truncated hierarchy carries a memory of the resolved fields through a convolution kernel \citep{gupta_generalized_2023}. In the specific case of Landau damping, the heat flux encodes the phase-mixing history of the distribution function. A trailing window of resolved moment fields provides a simple, discrete approximation to this memory. In practice we use $K=50$ steps at the training output cadence of $\Delta t = 0.1\,\omega_{\mathrm{p}}^{-1}$, i.e. a memory horizon of $5\,\omega_{\mathrm{p}}^{-1}$; at early times, before $K$ states are available, the window is left-padded with the corresponding equilibrium values ($n=1$, $p=1$, $u=E=\partial_x q = 0$).
 
We choose to have the network output the heat-flux gradient directly, rather than the heat flux itself, for two reasons. First, as the heat flux appears in the pressure equation only as a gradient, learning the gradient directly is a simpler training target and avoids the additional error introduced by approximating a further derivative through, e.g., finite differences. Second, and more subtly, this choice preserves the closure's ability to reproduce the net heating associated with Landau damping. Landau damping transfers energy from the wave to resonant particles, so we expect the spatially averaged pressure $\langle p\rangle = \tfrac{1}{L}\int p\,\mathrm{d}x$ to rise as the perturbation damps. On the periodic domain the heat-flux term contributes to $\tfrac{\mathrm{d}}{\mathrm{d}t}\langle p\rangle$ only through the spatial average $\langle \partial_x q\rangle$. Had the network instead produced a periodic field $q$ that we then differentiated, this average would vanish identically, $\int \partial_x q\,\mathrm{d}x = 0$, removing the heat-flux term as a channel for changing the mean pressure and leaving the closure unable to reproduce the heating. Outputting $\partial_x q$ directly places no such constraint on its spatial mean, allowing the learned closure to capture the net thermodynamic effect of the damping.

Time integration of the fluid equations (\ref{eq:continuity}, \ref{eq:momentum}, \ref{eq:pressure}) set out in Section~\ref{sec:background} is performed using a third-order Strong Stability Preserving Runge-Kutta method (SSPRK3). For a state $x$ evolving according to $\partial x / \partial t = f(x, t)$ with initial value $x_0$ at time $t = t_0$, a single SSPRK3 step of size $\Delta t$ produces
\begin{align}
    x^{(1)} &= x_0 + \Delta t \, f(x_0,\, t_0) \\
    x^{(2)} &= \tfrac{3}{4} x_0 + \tfrac{1}{4}
        \bigl(x^{(1)} + \Delta t \, f(x^{(1)},\, t_0 + \Delta t)\bigr) \\
    x^{(3)} &= \tfrac{1}{3} x_0 + \tfrac{2}{3}
        \bigl(x^{(2)} + \Delta t \, f(x^{(2)},\, t_0 + \tfrac{\Delta t}{2})\bigr),
\end{align}
where superscripts label the SSPRK3 substages and $x^{(3)}$ is the accepted update at $t_0 + \Delta t$.

The presence of the FNO closure in the time-derivative function $f$ requires care at each substage, because the closure is not a pointwise function of the instantaneous state but is conditioned on a trailing window of the moment history. Writing the per-slot channel bundle as
\begin{equation}
    \mathbf{m}_{t} \equiv
    \bigl(n_{t},\, u_{t},\, p_{t},\, E_{t},\, (\partial_x q)_{t}\bigr),
\end{equation}
the FNO takes as input the last $K$ slots of the history window,
\begin{equation}
    \mathbf{M}_{t} \equiv
    \bigl(\mathbf{m}_{t-(K-1)\Delta t},\, \dots,\, \mathbf{m}_{t}\bigr),
    \qquad
    (\partial_x q) = \mathrm{FNO}\bigl(\mathbf{M}_{t};\, \log A_{\mathrm{scale}}\bigr),
\end{equation}
together with the log-amplitude conditioning scalar $\log A_{\mathrm{scale}}$.
Here $K$ is the memory length. Denote by $\mathcal{F}_n$, $\mathcal{F}_u$, $\mathcal{F}_p$ the right-hand sides of (\ref{eq:continuity}, \ref{eq:momentum}, \ref{eq:pressure}) and by $\mathcal{P}$ the spectral Poisson solve returning $E$ from the density.

The subtlety introduced by the trailing window is that each provisional substage state must be inserted into the buffer before the FNO can be evaluated on it. We handle this with a single rolling slot: Stage~1 appends its predictor to the buffer, while Stages~2 and~3 overwrite that trailing slot, so at most one provisional entry is present at any time and, after Stage~3, the trailing slot holds the accepted state at $t_{i+1}$. The electric field is recomputed from Poisson at every substage, so the state fed to the FNO always satisfies the electrostatic constraint. Note also that the pressure update at each substage uses the closure produced at the \emph{previous} substage: Stage~1 uses the closure carried from $t_i$, and the FNO evaluated at the end of a substage supplies the closure consumed by the next. The explicit scheme advancing the accepted state $\{\mathbf{m}_{t_{i-K+1}}, \dots, \mathbf{m}_{t_i}\}$ by one step is:

\paragraph{Stage 1 (forward-Euler predictor).}
\begin{align}
    n^{(1)}          &= n_{t_i} + \Delta t\, \mathcal{F}_n\bigl(n_{t_i}, u_{t_i}\bigr) \\
    u^{(1)}          &= u_{t_i} + \Delta t\, \mathcal{F}_u\bigl(n_{t_i}, u_{t_i}, p_{t_i}, E_{t_i}\bigr) \\
    p^{(1)}          &= p_{t_i} + \Delta t\, \mathcal{F}_p\bigl(u_{t_i}, p_{t_i}, (\partial_x q)_{t_i}\bigr) \\
    E^{(1)}          &= \mathcal{P}\bigl(n^{(1)}\bigr) \\
    (\partial_x q)^{(1)} &= \mathrm{FNO}\bigl(\mathbf{M}_{t_i};\, \log A_{\mathrm{scale}}\bigr),
    \quad
    \mathbf{M}_{t_i} = \bigl(\mathbf{m}_{t_{i-K+1}}, \dots, \mathbf{m}_{t_i}\bigr).
\end{align}
Append $\mathbf{m}^{(1)} = (n^{(1)}, u^{(1)}, p^{(1)}, E^{(1)}, (\partial_x q)^{(1)})$
to the trailing memory window.

\paragraph{Stage 2.}
\begin{align}
    n^{(2)}          &= \tfrac{3}{4} n_{t_i} + \tfrac{1}{4}\bigl(n^{(1)} + \Delta t\, \mathcal{F}_n(n^{(1)}, u^{(1)})\bigr) \\
    u^{(2)}          &= \tfrac{3}{4} u_{t_i} + \tfrac{1}{4}\bigl(u^{(1)} + \Delta t\, \mathcal{F}_u(n^{(1)}, u^{(1)}, p^{(1)}, E^{(1)})\bigr) \\
    p^{(2)}          &= \tfrac{3}{4} p_{t_i} + \tfrac{1}{4}\bigl(p^{(1)} + \Delta t\, \mathcal{F}_p(u^{(1)}, p^{(1)}, (\partial_x q)^{(1)})\bigr) \\
    E^{(2)}          &= \mathcal{P}\bigl(n^{(2)}\bigr) \\
    (\partial_x q)^{(2)} &= \mathrm{FNO}\bigl(\mathbf{M}^{(1)};\, \log A_{\mathrm{scale}}\bigr),
    \quad
    \mathbf{M}^{(1)} = \bigl(\mathbf{m}_{t_{i-K+2}}, \dots, \mathbf{m}_{t_i}, \mathbf{m}^{(1)}\bigr).
\end{align}
Overwrite the trailing slot with $\mathbf{m}^{(2)}$.

\paragraph{Stage 3.}
\begin{align}
    n^{(3)}          &= \tfrac{1}{3} n_{t_i} + \tfrac{2}{3}\bigl(n^{(2)} + \Delta t\, \mathcal{F}_n(n^{(2)}, u^{(2)})\bigr) \\
    u^{(3)}          &= \tfrac{1}{3} u_{t_i} + \tfrac{2}{3}\bigl(u^{(2)} + \Delta t\, \mathcal{F}_u(n^{(2)}, u^{(2)}, p^{(2)}, E^{(2)})\bigr) \\
    p^{(3)}          &= \tfrac{1}{3} p_{t_i} + \tfrac{2}{3}\bigl(p^{(2)} + \Delta t\, \mathcal{F}_p(u^{(2)}, p^{(2)}, (\partial_x q)^{(2)})\bigr) \\
    E^{(3)}          &= \mathcal{P}\bigl(n^{(3)}\bigr) \\
    (\partial_x q)^{(3)} &= \mathrm{FNO}\bigl(\mathbf{M}^{(2)};\, \log A_{\mathrm{scale}}\bigr),
    \quad
    \mathbf{M}^{(2)} = \bigl(\mathbf{m}_{t_{i-K+2}}, \dots, \mathbf{m}_{t_i}, \mathbf{m}^{(2)}\bigr).
\end{align}
Overwrite the trailing slot with $\mathbf{m}^{(3)}$ and accept $\mathbf{m}_{t_{i+1}} = \mathbf{m}^{(3)}$.

This scheme was chosen as the most straightforward implementation of the online closure. Its main drawback is that the trained FNO is tightly coupled to this particular numerical method, which may limit its portability to other fluid codes; a more solver-independent implementation is left to future work.

Spatial derivatives in the fluid equations are computed via second-order central finite differences on the 128-point grid, utilising the periodic boundary conditions. The fluid solver shares the same spatial domain, $L = 2 \upi / k$, as the kinetic simulation. The entire differentiable fluid solver, including the SSPRK3 time integration, the spectral Poisson solve and the finite-difference spatial operators, is implemented in PyTorch \citep{paszke_pytorch_2019}, so that the full fluid rollout supports reverse-mode automatic differentiation and gradients can be propagated back through it to the FNO parameters during online training.

\subsection{Fourier Neural Operator}
\label{sec:fno}

We use a Fourier Neural Operator (FNO) as the architecture for our closure function \citep{li_fourier_2021}, implemented using the \texttt{FNO} class of the \texttt{neuraloperator} Python package \citep{kossaifi_neural_2024}. The closure network maps the trailing window of resolved moment fields to the heat-flux gradient one output step ahead. Each of the five resolved fields $\{n,u,p,E,\partial_x q\}$ is supplied over the $K=50$ most recent timesteps on the $128$-point spatial grid, with the density and pressure entered as perturbations about their unit equilibrium ($n-1$ and $p-1$), and a single constant channel carries the amplitude conditioning variable $\log A_{\mathrm{scale}}$, giving $5K+1 = 251$ input channels of length $128$. The network outputs a single channel, the predicted $\partial_x q$ at the next step, also of length $128$. The FNO comprises $8$ Fourier layers, each retaining the lowest $32$ Fourier modes and using $64$ hidden channels. Because Landau damping is periodic and dominated by low-wavenumber structure, this truncated-mode architecture is well matched to the problem \citep{li_fourier_2021}; the periodic geometry likewise makes the Fourier layers a natural choice. The online training method is in principle independent of the closure architecture, and a systematic comparison of the FNO against alternative architectures, such as other operator-learning approaches \citep{lu_learning_2021}, trained under the same online procedure is a worthwhile avenue for further study, which we leave to future work. These architectural choices are common to every closure trained in this work, the single-amplitude linear and nonlinear closures and the multi-amplitude closure, and are summarised in Table~\ref{tab:fno_arch}.

\begin{table}
  \centering
  \begin{tabular}{lc}
    \toprule
    Hyperparameter & Value \\
    \midrule
    Fourier layers & 8 \\
    Retained Fourier modes (per layer) & 32 \\
    Hidden channels & 64 \\
    Input channels ($5K+1$) & 251 \\
    Output channels & 1 \\
    Activation function & GELU \\
    Block precision & full \\
    \bottomrule
  \end{tabular}
  \caption{FNO closure architecture, common to all trained models (the single-amplitude linear and nonlinear closures and the multi-amplitude closure). The number of input channels follows from the five resolved fields $\{n,u,p,E,\partial_x q\}$ supplied over the $K=50$-step memory window, together with the single spatially constant $\log A_{\mathrm{scale}}$ conditioning channel, giving $5K+1=251$. The architecture is implemented using the \texttt{FNO} class of the \texttt{neuraloperator} library \citep{kossaifi_neural_2024}.}
  \label{tab:fno_arch}
\end{table}

Supplying a history of the fields, rather than only the instantaneous state, allows the closure to represent the non-Markovian dependence discussed in Section~\ref{sec:fluid_model}. During a fluid rollout the memory window is populated from the fluid trajectory itself: at each SSPRK3 substage the FNO is evaluated on the current and recent resolved states, so that the closure is consistent with the same finite-memory map it was trained under.
 
A practical consideration that proved central to obtaining good performance is the scaling of the FNO inputs. The fluid moments span very different magnitudes: the density and the electric field are typically of order unity in our normalisation, while the heat flux can be many orders of magnitude smaller, particularly in the linear regime; moreover the magnitude of each moment varies significantly over the course of a single simulation, and across simulations by the several orders of magnitude that separate the linear and nonlinear regimes. We address this through two scaling steps applied to the FNO inputs and outputs.

First, we normalise the perturbation fields by a characteristic amplitude computed from the most recent density field in the input window,
\begin{equation}
  A_{\mathrm{scale}}
  = \left(\frac{1}{N}\sum_{x} \bigl(n(x)-1\bigr)^{2}\right)^{\!1/2},
  \label{eq:ascale}
\end{equation}
where $N$ is the number of grid points in $x$ and $n(x)$ is the density at the most recent timestep in the memory window; $A_{\mathrm{scale}}$ is therefore the root-mean-square amplitude of the density perturbation, and is floored at a small value ($10^{-6}$) to remain well defined at exact equilibrium. Here the perturbation fields are the deviations of each moment from its spatially uniform equilibrium: the density and pressure are supplied to the network as $n-1$ and $p-1$, their normalised background values both being fixed at unity, whereas the velocity $u$, electric field $E$ and heat-flux gradient $\partial_x q$ vanish at equilibrium and are supplied directly. Each perturbation field is divided by $A_{\mathrm{scale}}$ before being supplied to the network, and the output $\partial_x q$ is multiplied by $A_{\mathrm{scale}}$ to restore physical units. Normalising in this way lets the network operate on order-unity fields irrespective of the underlying amplitude regime, so that it learns the \emph{shape} of the closure relation rather than its overall scale; without this step, training would be biased towards the larger-amplitude cases at the expense of the linear regime, where the perturbations are vanishingly small. Because normalisation removes the amplitude, we reintroduce it explicitly as a conditioning input by appending $\log A_{\mathrm{scale}}$ as an additional, spatially constant channel. This tells the network which amplitude regime it is operating in and allows a single model to modulate its behaviour continuously between the linear (small $A_{\mathrm{scale}}$) and nonlinear (large $A_{\mathrm{scale}}$) limits, and to interpolate between the amplitude scales seen in training.

Second, we apply an inverse hyperbolic sine ($\mathrm{arcsinh}$) transformation, which acts approximately linearly on small inputs and approximately logarithmically on large inputs. This compresses the dynamic range of the inputs while preserving the structure of the small scales, which is crucial in the linear regime where perturbation amplitudes become very small. Although input scaling is not generally given prominent treatment in the existing machine-learning plasma closure literature, in our experience it is one of the most consequential design choices for producing closures that perform well across a wide range of perturbation amplitudes.

\subsection{Online Training Method}
\label{sec:online_method}

The online training procedure operates as follows. At each training iteration, a batch of kinetic snapshots is sampled from the training set; each snapshot provides the initial conditions for a short rollout of the closed fluid simulation, in which the FNO replaces the heat-flux gradient term as described in Section~\ref{sec:fluid_model}. The fluid rollout is integrated for a chosen number of timesteps, with the FNO evaluated at each substage of the SSPRK3 update. The resulting fluid trajectory is compared with the corresponding kinetic trajectory at the same time points, and the loss is computed as the relative $L^{2}$ norm between the two over all rolled-out timesteps and all evolved moments. Because the entire fluid solver is implemented in PyTorch \citep{paszke_pytorch_2019}, a differentiable framework, gradients of the loss with respect to the FNO parameters can be computed through the rollout via automatic differentiation, and standard stochastic gradient methods can therefore be used to update the parameters.

This procedure differs fundamentally from offline training in that the loss never references the kinetic heat flux directly. The FNO is trained only against the trajectories of the lower-order moments and is free to converge to whichever heat-flux profile produces the best agreement with those trajectories under the specific fluid solver in use. This is, in general, not expected to be identical to the heat flux directly calculated from the kinetic simulation, due to differences in numerical scheme, discretisation and time-stepping between the two simulation methods. We also interpret this discrepancy between `kinetic' heat flux and the `effective fluid' heat flux as due to the loss of information when truncating the moment hierarchy. The kinetic model implicitly includes all higher-order moments, whereas the fluid model only retains moments up to the heat flux, and thus the `effective heat' flux that reproduces kinetic effects in the fluid model will need to compensate for the lack of higher-order moments, leading to the pointwise differences between the learned heat flux and the kinetic heat flux.

For the dataset used in single-amplitude experiments, the train/test split is taken chronologically; the earlier portion of each simulation provides the training snapshots and the later portion provides the test snapshots. For the generalisation experiments described in Section~\ref{sec:generalisation}, the split is instead taken across initial perturbation amplitudes, as described in that section.

Practical implementation requires care with memory and compute. Each fluid rollout requires storing intermediate states for backpropagation, and the memory footprint scales with the number of rollout steps; longer rollouts therefore force smaller batch sizes and require additional accelerator memory. Tuning these hyperparameters involves a trade-off: shorter rollouts permit faster training and larger batches but provide a weaker training signal for the long-time behaviour of the closure, while longer rollouts improve the final performance and stability at substantial computational cost. All training conducted for this investigation was performed on a single NVIDIA A100 GPU with $40 \, \mathrm{GB}$ of VRAM. In each epoch the full set of training snapshots is sampled once, each snapshot seeding a separate 100-step fluid rollout, with the batch size chosen dynamically to fill the available GPU memory; on this hardware that permitted a batch size of 50. For each model discussed here, training was performed with a rollout length of 100 fluid timesteps at a timestep length of $\Delta t = 0.1 \omega_{\mathrm{p}}^{-1}$, which we found to balance well between ensuring the stability of the trained closure, while remaining computationally viable. In future work, it will be informative to conduct online training with different lengths of $\Delta t$, specifically investigating if ML closures can reproduce kinetic effects even when trained and/or tested with fluid-like timescales. The training hyperparameters, common to all models, are summarised in Table~\ref{tab:train_hyper}.

\begin{table}
  \centering
  \begin{tabular}{lc}
    \toprule
    Hyperparameter & Value \\
    \midrule
    Spatial grid points, $N_G$        & 128 \\
    Fluid timestep, $\Delta t$        & $0.1\,\omega_{\mathrm{p}}^{-1}$ \\
    Rollout length                    & 100 steps ($10\,\omega_{\mathrm{p}}^{-1}$) \\
    Memory window, $K$                & 50 steps ($5\,\omega_{\mathrm{p}}^{-1}$) \\
    Loss                              & relative $L^{2}$ (all moments except $\partial_x q$, all rollout steps) \\
    Optimiser                         & AdamW \\
    Learning rate (initial)           & $10^{-3}$ \\
    Learning-rate schedule            & cosine annealing ($T_{\max}=1000$ epochs, $\eta_{\min}=0$) \\
    Training epochs                   & 1000 \\
    Batch size                        & 50 \\
    Gradient clipping                 & none \\
    \bottomrule
  \end{tabular}
  \caption{Training hyperparameters, common to all trained models. The batch size is selected dynamically to make use of the available GPU memory; the value of 50 corresponds to the 40\,GB A100 used here. In each epoch the entire set of training snapshots is sampled once, each snapshot providing the initial condition for a 100-step fluid rollout. The models differ only in their training data: the single-amplitude closures are trained on one initial amplitude with a chronological train/test split, whereas the multi-amplitude closure is trained on the amplitudes defined in Section~\ref{sec:generalisation} with the split taken across amplitudes.}
  \label{tab:train_hyper}
\end{table}

\subsection{Offline Training for Comparison}
\label{sec:offline}

To isolate the effect of online training, we compare against an otherwise identical closure trained offline (\textit{a priori}). The offline model is deliberately matched to the online model in every respect except the training objective: it uses the same FNO architecture and hyperparameters (Section~\ref{sec:fno}), the same input construction, including the $K=50$-step trailing memory window and the $\log A_{\mathrm{scale}}$ conditioning channel, the same input scaling, and the same relative $L^2$ loss and optimiser settings. The single difference is the origin of the data on which the loss is evaluated. In the offline case the memory windows are sampled directly from the kinetic simulation trajectories, and the network is trained to regress the kinetic heat-flux gradient, $\partial_x q_{\mathrm{kin}}$, computed from those same kinetic snapshots; the loss never involves a fluid rollout. This is in contrast to the online case, in which each memory window is generated by the closed fluid solver during the rollout and the loss is evaluated on the resulting fluid trajectory (Section~\ref{sec:online_method}).

Both models are then evaluated identically. Testing is performed online in every case: the trained closure is embedded in the fluid solver, initialised from the same conditions as the corresponding kinetic run, and integrated over the full simulation duration, with the resulting trajectory compared against the kinetic ground truth. All comparisons between the online- and offline-trained closures reported in Section~\ref{sec:results} therefore differ only in how the closure was trained, not in how it is deployed or assessed.

We stress that this matched comparison is intended to isolate the effect of the training method, and not to argue that offline training cannot produce competent closures for this problem. Offline-trained closures have been shown to work well in this setting: \citet{huang_machine-learning_2025}, for example, obtained an offline-trained FNO closure that reproduced both the linear and nonlinear regimes of one-dimensional Landau damping. The offline results reported here should accordingly be read as specific to our matched configuration rather than as a general limitation of offline training, since design choices left fixed for the comparison can materially affect offline performance. The temporal sampling of the training data is one such choice: \citet{huang_machine-learning_2025} sampled their training snapshots at a considerably finer timestep than the output cadence used here, which provides the offline regression with a denser and more informative target and may partly explain the stronger offline performance they report. A systematic study of how data sampling and related choices affect offline relative to online training is beyond our present scope; our narrower claim is that, under identical hyperparameters and training data, we find online training to be the more robust of the two.

\subsection{Hammett-Perkins closure benchmark}
\label{sec:hp}

As a second, non-learned reference we deploy the analytic Hammett-Perkins closure (\ref{eq:hp}) in the same fluid solver. Unlike the FNO, the Hammett-Perkins closure is a memoryless algebraic map from the instantaneous resolved state to the heat-flux gradient; we therefore evaluate it inline at every SSPRK3 stage, on the stage's own $(n,p)$, rather than under the trailing-window substep convention used for the FNO (Section~\ref{sec:fluid_model}). The heat flux is formed from the $\mathrm{i} \, \mathrm{sgn}(k)$ multiplier by FFT and differentiated with the same finite-difference operator as the rest of the solver, so the discretisation is consistent across the two.

The coefficient in (\ref{eq:hp}) is fixed by the linear kinetic response, but the three-pole Pad\'{e} approximation underlying it reproduces the exact damping rate only approximately at any single wavenumber. At our $k = 0.35$ the canonical coefficient $\sqrt{8/\upi} \approx 1.60$ over-damps the mode; we therefore calibrate the coefficient to match the kinetic linear damping rate, obtaining $\chi = \sqrt{4.8/\upi} \approx 1.24$, $22\%$ below the canonical value. This offset is consistent with the known mode-dependent accuracy of the three-pole closure, which fits the response function across all phase velocities rather than reproducing the exact rate at one. Calibrating in this way gives the Hammett-Perkins closure its best-case linear performance, so that any departure in the nonlinear regime reflects the structural limitations of the closure rather than a sub-optimal coefficient.

\section{Results}
\label{sec:results}

In this section we present the performance of the online-trained FNO closure across three experimental settings. Section~\ref{sec:linear} examines the case of linear Landau damping, Section~\ref{sec:nonlinear} focuses on the nonlinear case, and Section~\ref{sec:generalisation} examines a single FNO trained across multiple initial perturbation amplitudes and tested on amplitudes both within and outside the training set. In each setting we benchmark the online-trained FNO against the same FNO architecture trained offline against the kinetic heat flux, and where appropriate against the analytic Hammett-Perkins closure (Section~\ref{sec:hp}).

\subsection{Linear Landau Damping}
\label{sec:linear}

\begin{figure}
  \centering
  \begin{subfigure}{.49\textwidth}
    \includegraphics[width=\linewidth]{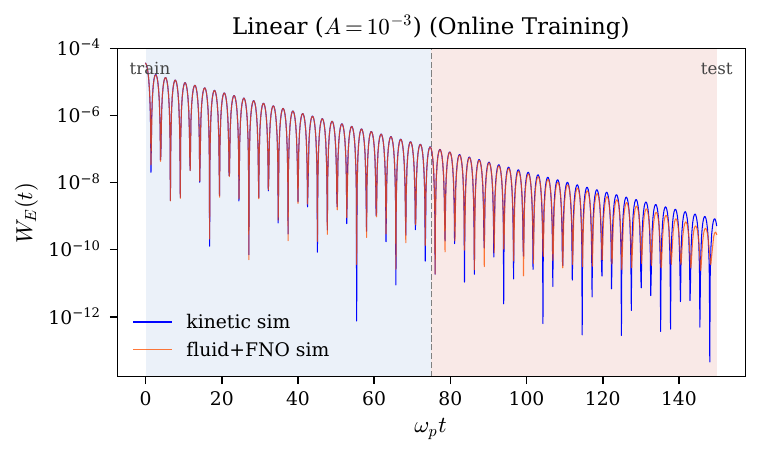}
    \caption{Online-trained closure.}\label{fig:online_linear_WE}
  \end{subfigure}\hfill
  \begin{subfigure}{.49\textwidth}
    \includegraphics[width=\linewidth]{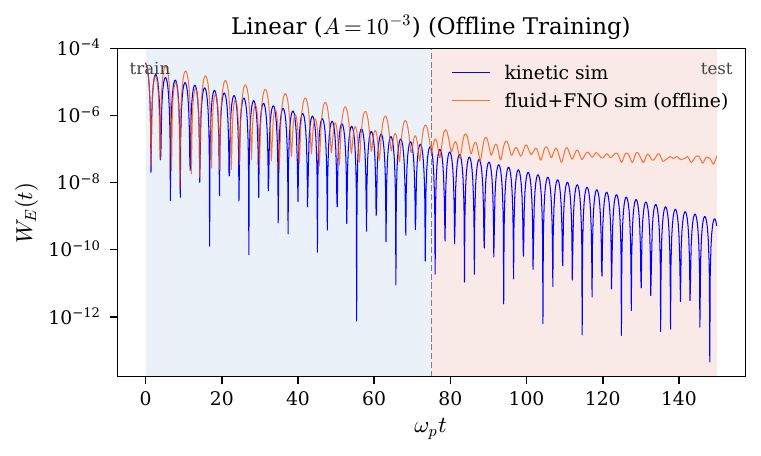}
    \caption{Offline-trained closure.}\label{fig:offline_linear_WE}
  \end{subfigure}

  \vspace{\baselineskip}
  \begin{subfigure}{.49\textwidth}
    \includegraphics[width=\linewidth]{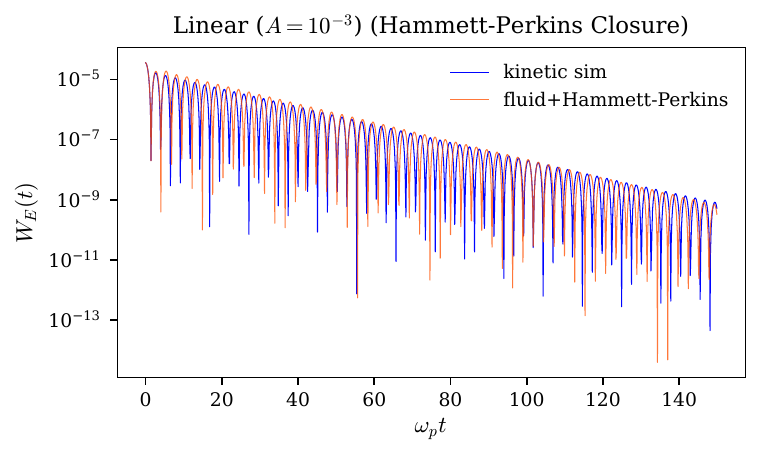}
    \caption{Hammett-Perkins closure.}\label{fig:hp_linear_WE}
  \end{subfigure}
  \caption{Electric-field energy $W_E(t)$ for the linear case ($A=10^{-3}$): kinetic ground truth against the fluid simulation closed with (a)~the online-trained FNO, (b)~the offline-trained FNO and (c)~the analytic Hammett-Perkins closure ($\chi=\sqrt{4.8/\upi}$). All panels share the same axes and initial conditions; shaded regions in (a) and (b) mark the FNO training window and do not apply to the untrained Hammett-Perkins closure in (c).}
  \label{fig:linear_WE}
\end{figure}

For the linear case we use an initial perturbation amplitude of $A = 10^{-3}$, which produces electric-field perturbations small enough that nonlinear coupling between modes remains negligible over the full simulation duration. Figure~\ref{fig:linear_WE} shows the electric-field energy $W_E = \tfrac{1}{2} \int E^2 \, \mathrm{d}x$ as a function of time for the kinetic simulation, the fluid simulation closed with the online-trained FNO (Figure~\ref{fig:online_linear_WE}) and the fluid simulation closed with the offline-trained FNO (Figure~\ref{fig:offline_linear_WE}). In both cases the fluid simulation is integrated over the full simulation time, $0<t<150\,\omega_{\mathrm{p}}^{-1}$ and uses the same initial conditions as the kinetic ``ground truth''. The shaded regions denoted with `train' and `test' mark the portions of the kinetic trajectory that were and were not sampled during training, the split falling at $t=75\,\omega_{\mathrm{p}}^{-1}$.

The online-trained closure reproduces the linear damping in detail (Figure~\ref{fig:online_linear_WE}). Throughout the training window the fluid $W_E(t)$ tracks the kinetic curve peak-for-peak, matching both the exponential decay envelope and the phase of the oscillation. We extract the damping rate by fitting a straight line to the peaks of $\log W_E$ over the training and test windows separately; the resulting rates are summarised in Table~\ref{tab:single_sim_gamma}. In the training window the fitted rate agrees with the kinetic value to within $0.2\%$ ($\gamma_\mathrm{FNO}=-3.43\times10^{-2}$ against $\gamma_\mathrm{kin}=-3.43\times10^{-2}$), confirming that the online closure recovers the correct Landau rate rather than merely a stable decay. This fidelity is reflected across all moments in Table~\ref{tab:single_sim_L2}: within the training window every evolved field is reproduced to a time-averaged relative $L^2$ error at or below the $10^{-2}$ level, with the density and pressure errors as low as $\sim10^{-6}$.

Beyond the training window the two trajectories begin to separate slowly. The fitted test-window damping rate departs from the kinetic value by $\approx18\%$ (Table~\ref{tab:single_sim_gamma}), and the moment errors grow accordingly, with the electric-field and velocity errors rising from the $10^{-3}$ level in training to $\sim10^{-1}$ in test while the density and pressure errors remain at the $10^{-6}$ level (Table~\ref{tab:single_sim_L2}). Table~\ref{tab:single_sim_divtime} quantifies the onset of this drift: the relative electric-field error crosses $10\%$ almost immediately on leaving the training window ($t_\mathrm{div}\approx1.4\,\omega_{\mathrm{p}}^{-1}$ after the split), but takes considerably longer to reach $50\%$ and $100\%$ ($45$ and $55\,\omega_{\mathrm{p}}^{-1}$ respectively), so the closure degrades gracefully rather than destabilising. The divergence is confined to the field variables that carry the oscillation phase: the deep nulls of $W_E$ fall out of phase and the FNO peaks drop slightly below the kinetic peaks at late times, while the smoothly varying density and pressure remain accurate.

\begin{figure}
    \centering
    \includegraphics[width=1\linewidth]{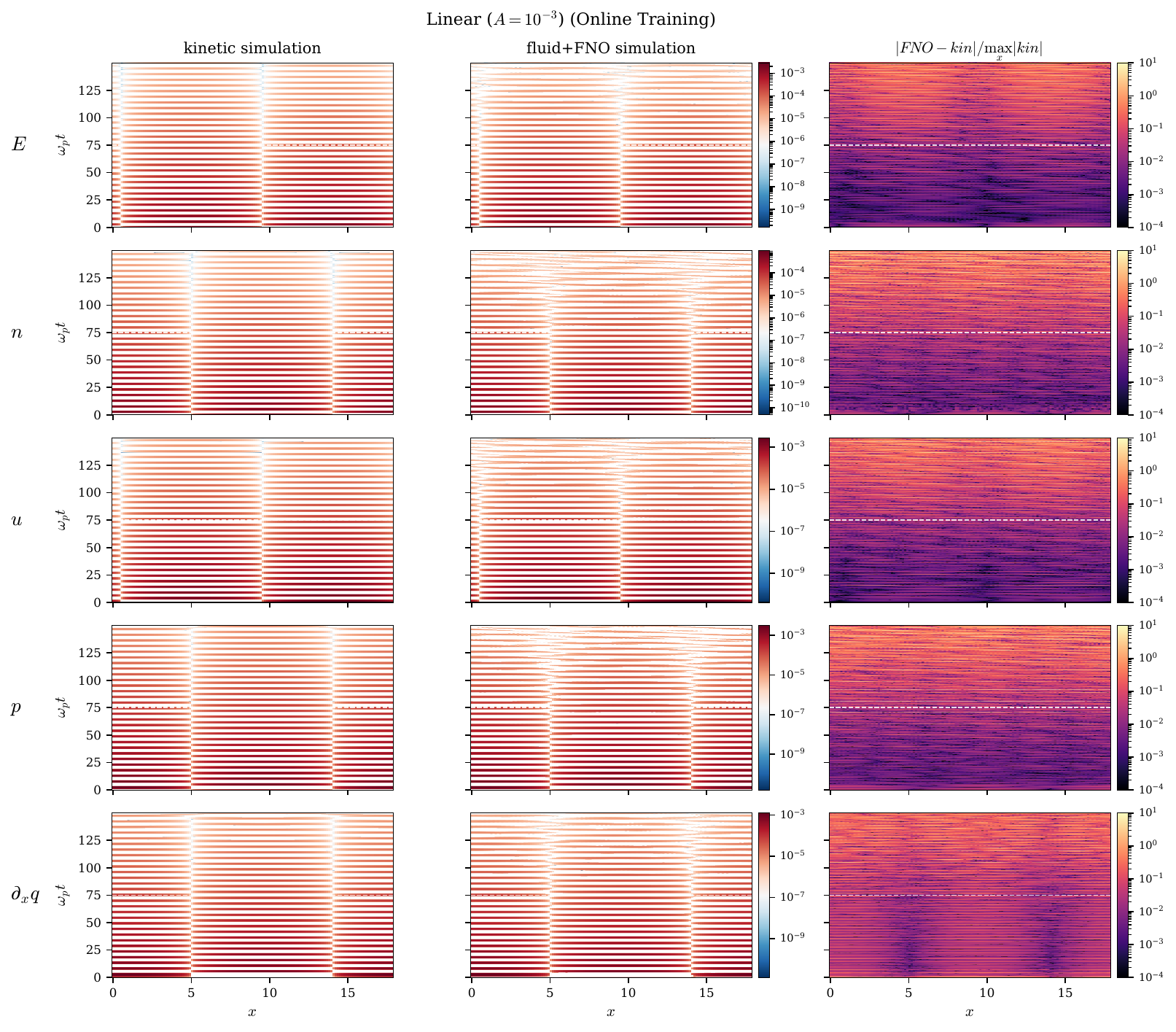}
    \caption{Space-time evolution of the evolved moments for the linear single-amplitude case ($A=10^{-3}$, online-trained closure). Each row corresponds to one field ($E$, $n$, $u$, $p$, $\partial_x q$); the first column shows the kinetic ground truth, the second the fluid+FNO simulation, and the third the pointwise relative error $|\mathrm{FNO}-\mathrm{kin}|/\max_x|\mathrm{kin}|$. Time runs vertically ($0$-$150\,\omega_{\mathrm{p}}^{-1}$) and space horizontally over one perturbation wavelength $L=2\upi/k$; colour scales are logarithmic and the horizontal dashed line marks the $t=75\,\omega_{\mathrm{p}}^{-1}$ train/test split. The kinetic and fluid columns are visually indistinguishable; the error column is small and spatially uniform through the training window and grows only gradually into the test window, with the late-time increase concentrated in $E$, $n$ and $u$ as the deeply damped signal approaches the numerical noise floor of the solver.}
    \label{fig:linear_contour}
\end{figure}

This late-time drift is expected. By the end of the run the linear signal has damped by many orders of magnitude below its initial value (Figure~\ref{fig:online_linear_WE}), into a regime the closure never saw during training: the chronological split means the smallest-amplitude, most deeply damped states lie entirely in the test window, and small phase errors accumulate over a rollout far longer than the $100$-step ($10\,\omega_{\mathrm{p}}^{-1}$) window used in training. Figure~\ref{fig:linear_contour} shows the relative error in each evolved fluid moment between the online-tested fluid simulation and the kinetic ground truth as a function of space and time, and confirms this picture: the error is small and spatially uniform through the training window, and grows only gradually after the split, remaining well controlled across the domain.

The offline-trained closure, matched to the online model in architecture, hyperparameters and training data (Section~\ref{sec:offline}), does not reproduce the linear dynamics under the same matched configuration (Figure~\ref{fig:offline_linear_WE}). Rather than following the oscillatory decay, it produces a smooth $W_E(t)$ that damps far too slowly and misses the oscillation structure entirely, sitting orders of magnitude above the kinetic minima by the end of the run. As discussed in Section~\ref{sec:offline}, this behaviour is specific to the matched comparison and is sensitive to the temporal sampling of the offline training data; it is not a claim that offline training cannot succeed on this problem.

The analytic Hammett-Perkins closure reproduces the linear damping well (Figure~\ref{fig:hp_linear_WE}), as expected of a closure constructed to capture precisely this regime. With the calibrated coefficient the Hammett-Perkins fluid $W_E(t)$ tracks the kinetic decay envelope and rate across the full run, with a fitted damping rate $\gamma_{\mathrm{HP}} = -3.52 \times 10^{-2}$ (Table~\ref{tab:single_sim_gamma}), matching the kinetic rate by construction. A closer look shows, however, that the calibrated closure reproduces the damping rate but not the oscillation frequency, with the peaks of $W_E$ drifting progressively out of alignment with the kinetic trace, a $\approx 5 \%$ error in the real oscillation frequency ($\omega_{\mathrm{r}} = 1.16$ against the kinetic 1.22) even as the decay envelope is matched. This is expected of a single-coefficient closure. Calibrating $\chi$ fixes the imaginary part of the complex frequency, leaving the real part set by the remaining fluid dispersion, which is not independently matched to the kinetic value. Because the online FNO and the Hammett-Perkins closure are deployed in the same fluid solver, the discrepancy cannot be attributed to the fluid discretisation, the online closure reproduces the kinetic $W_E$ peak-for-peak (prior to reaching times unseen during training), matching both the damping envelope and the oscillation phase, whereas Hammett-Perkins matches only the envelope. Even in the linear regime, then, the learned closure captures the full linear dispersion that the analytic closure only partially reproduces.

\begin{table}
  \centering
  \begin{tabular}{lcc}
    \toprule
     & \multicolumn{2}{c}{Linear ($A = 10^{-3}$)} \\
    \cmidrule(lr){2-3}
     & Train ($t<75$) & Test ($t\geq75$) \\
    \midrule
    $\gamma_\mathrm{kin}$ & $-3.43\times10^{-2}$ & $-3.30\times10^{-2}$ \\
    $\gamma_\mathrm{FNO (online)}$ & $-3.43\times10^{-2}$ & $-3.91\times10^{-2}$ \\
    $\gamma_\mathrm{HP}$     & $-3.53 \times 10^{-2}$ & $-3.54 \times 10^{-2}$ \\
    Rel.\ err.\ (FNO (online)) & $1.56\times10^{-3}$ & $1.82\times10^{-1}$ \\
    Rel.\ err.\ (HP)         & $2.69 \times 10^{-2}$ & $6.94 \times 10^{-2}$ \\
    \bottomrule
  \end{tabular}
  \caption{
    Damping rates fitted to the peaks of $\log W_E(t)$ over the in-training (Train) and out-of-training (Test) windows, for the linear single-simulation case, for both the online-trained FNO and the Hammett-Perkins closures. The relative error is $|\gamma_\mathrm{i} - \gamma_\mathrm{kin}|/|\gamma_\mathrm{kin}|$.}
  \label{tab:single_sim_gamma}
\end{table}

\subsection{Non-Linear Landau Damping}
\label{sec:nonlinear}

\begin{figure}
  \centering
  \begin{subfigure}{.49\textwidth}
    \includegraphics[width=\linewidth]{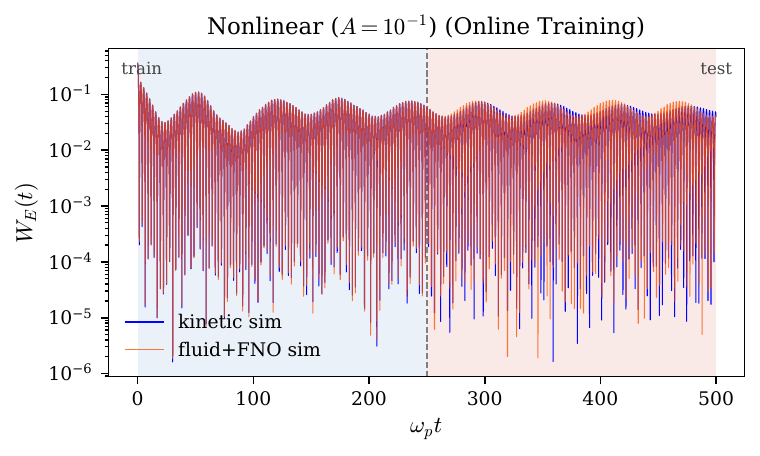}
    \caption{Online-trained closure.}\label{fig:online_nonlinear_WE}
  \end{subfigure}\hfill
  \begin{subfigure}{.49\textwidth}
    \includegraphics[width=\linewidth]{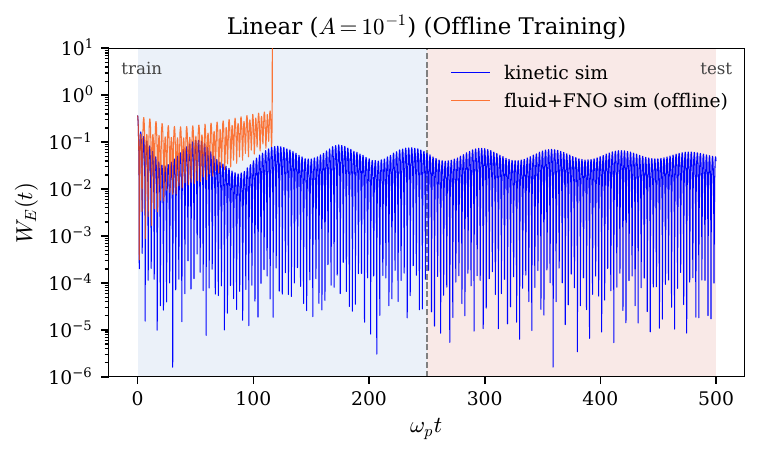}
    \caption{Offline-trained closure.}\label{fig:offline_nonlinear_WE}
  \end{subfigure}

  \vspace{\baselineskip}
  \begin{subfigure}{.49\textwidth}
    \includegraphics[width=\linewidth]{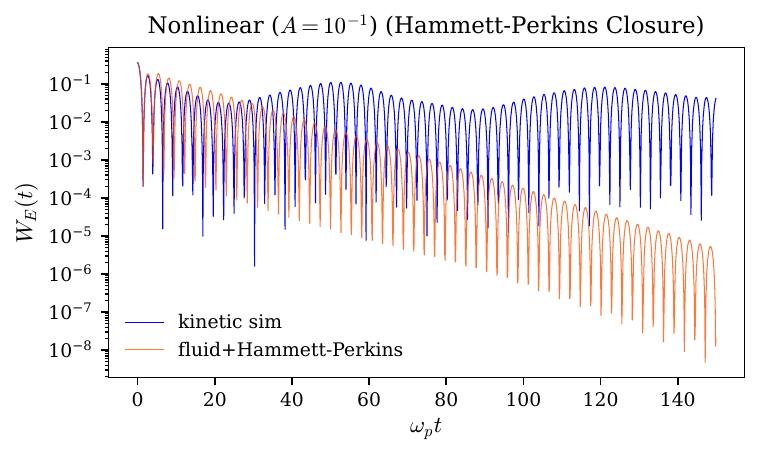}
    \caption{Hammett-Perkins closure.}\label{fig:hp_nonlinear_WE}
  \end{subfigure}
  \caption{Electric-field energy $W_E(t)$ for the nonlinear case ($A=10^{-1}$): kinetic ground truth against the fluid simulation closed with (a)~the online-trained FNO, (b)~the offline-trained FNO and (c)~the analytic Hammett-Perkins closure ($\chi=\sqrt{4.8/\upi}$). All panels share the initial conditions; shaded regions in (a) and (b) mark the FNO training window and do not apply to the untrained Hammett-Perkins closure in (c). Panel (c) is restricted to $t < 15 \, \omega_{\mathrm{p}}^{-1}$ for demonstrative purposes.}
  \label{fig:nonlinear_WE}
\end{figure}

For the nonlinear case we increase the initial perturbation amplitude to $A = 10^{-1}$, large enough that the system enters the trapping regime: the electric field undergoes initial damping followed by saturation and a subsequent slow modulation at the bounce frequency. Figure~\ref{fig:nonlinear_WE} shows $W_E(t)$ for the kinetic simulation, the fluid simulation closed with the online-trained FNO (Figure~\ref{fig:online_nonlinear_WE}) and the fluid simulation closed with the offline-trained FNO (Figure~\ref{fig:offline_nonlinear_WE}).

The online-trained closure reproduces the full nonlinear evolution across the entire $500\,\omega_{\mathrm{p}}^{-1}$ run (Figure~\ref{fig:online_nonlinear_WE}). It captures the initial damping phase, the level at which $W_E$ saturates, and the period and amplitude of the bounce oscillation that modulates the saturated envelope, and it does so not only within the training window but throughout the test window ($t>250\,\omega_{\mathrm{p}}^{-1}$), where the envelope and bounce phase continue to track the kinetic reference. The resolved moments confirm this: Table~\ref{tab:single_sim_L2} shows the density and pressure errors remaining at the $10^{-3}$-$10^{-2}$ level in the test window and the electric-field and velocity errors at $\sim10^{-1}$, an order of magnitude or more below the corresponding late-time errors of the linear case despite the far longer integration. The nonlinear closure is also markedly more robust to leaving the training window than the linear one: Table~\ref{tab:single_sim_divtime} shows the electric-field error reaching $50\%$ and $100\%$ only after $207$ and $270\,\omega_{\mathrm{p}}^{-1}$ respectively, against $45$ and $55\,\omega_{\mathrm{p}}^{-1}$ in the linear case, so the $100\%$ threshold is not crossed within the simulated window at all. The saturated field energy remains an $O(1)$ signal well above both the numerical noise floor and the amplitude of any recurrence revival, and the bounce dynamics are quasi-periodic and therefore well represented within the sampled training window, both of which help the closure generalise to later times.

\begin{figure}
    \centering
    \includegraphics[width=1\linewidth]{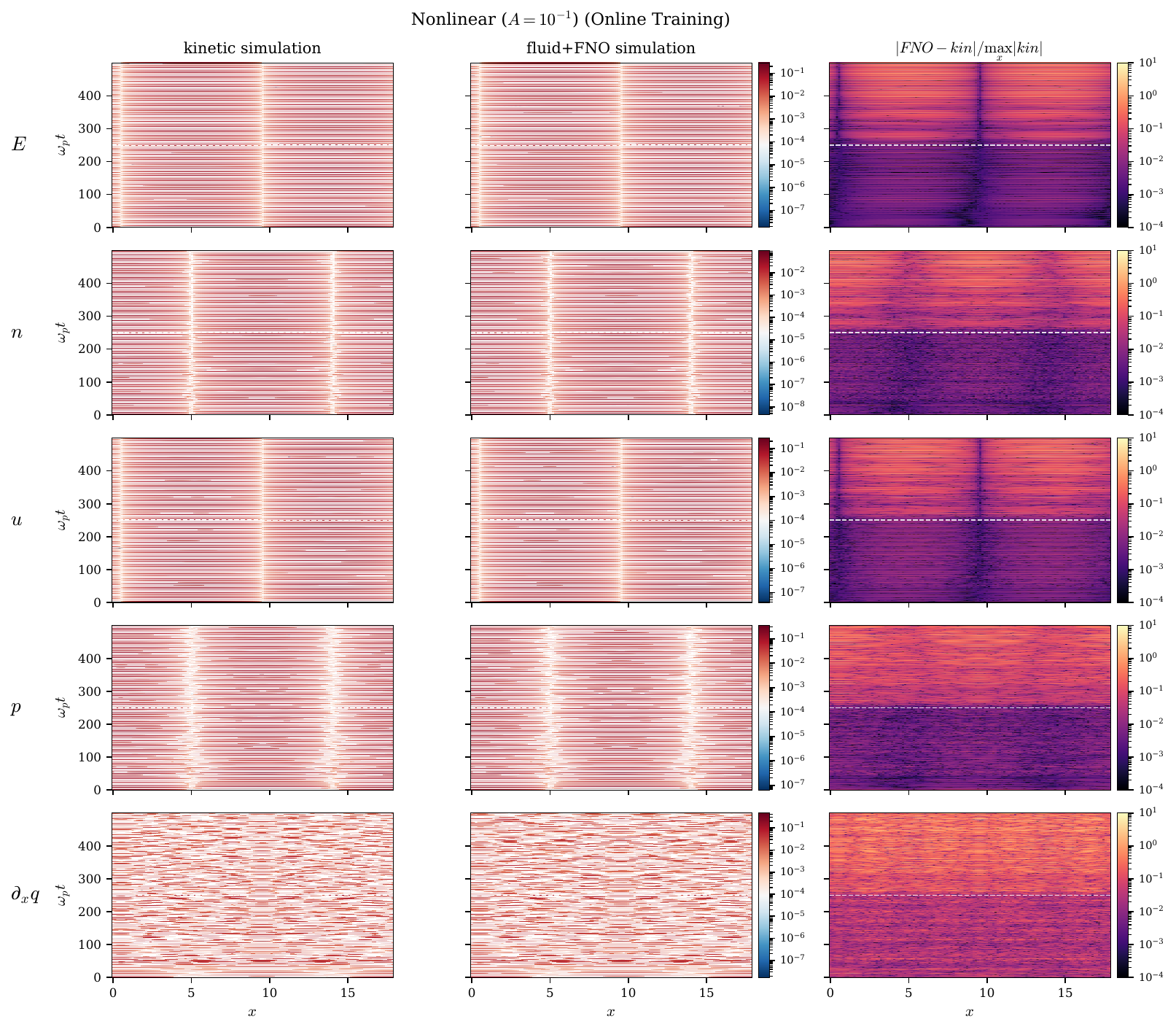}
    \caption{Space-time evolution of the evolved moments for the nonlinear single-amplitude case ($A=10^{-1}$, online-trained closure), in the same layout as Figure~\ref{fig:linear_contour}. Rows are the fields $E$, $n$, $u$, $p$, $\partial_x q$; columns are the kinetic ground truth, the fluid+FNO simulation, and the pointwise relative error $|\mathrm{FNO}-\mathrm{kin}|/\max_x|\mathrm{kin}|$. Time runs vertically ($0$-$500\,\omega_{\mathrm{p}}^{-1}$) and space over one wavelength $L=2\upi/k$; colour scales are logarithmic and the dashed line marks the $t=250\,\omega_{\mathrm{p}}^{-1}$ train/test split. The errors in $E$, $n$, $u$ and $p$ remain bounded over the full run, their spatial structure tracking the trapping-generated phase-space structure that the truncated moments cannot resolve, while the $\partial_x q$ error is large and broadly distributed, reflecting the departure of the learned effective heat flux from the kinetic one (Section~\ref{sec:heatflux}).}
    \label{fig:nonlinear_contour}
\end{figure}

The one field that departs substantially from the kinetic reference is the heat-flux gradient itself. The time-averaged $\partial_x q$ error grows from $\sim10^{-1}$ in training to order unity in the test window (Table~\ref{tab:single_sim_L2}), even as the observable fluid dynamics remain well reproduced. Figure~\ref{fig:nonlinear_contour} shows the relative error in each fluid moment over space and time: the errors in $E$, $n$, $u$ and $p$ remain bounded and moderate over the full duration, with their spatial structure reflecting the phase-space structures generated by particle trapping that the truncated moments cannot fully resolve, whereas the $\partial_x q$ error is large and broadly distributed across the domain. That the closure reproduces the resolved-moment dynamics while producing a heat flux that diverges pointwise from the kinetic one is the central observation we return to in Section~\ref{sec:heatflux}.

The offline-trained closure again fails under the matched configuration (Figure~\ref{fig:offline_nonlinear_WE}). It does not settle to the correct saturation level, instead sitting well above the kinetic envelope and drifting slowly upward rather than saturating, and it does not reproduce the oscillatory structure of the field energy. As in the linear case, and as discussed in Section~\ref{sec:offline}, this reflects the matched offline configuration rather than a general limitation of offline training.

The Hammett-Perkins closure cannot follow the nonlinear evolution (Figure~\ref{fig:hp_nonlinear_WE}). After the initial damping it fails to reproduce the saturation and the bounce modulation of the trapped state, since trapping is a nonlinear kinetic effect lying outside the linear theory from which the closure is derived, and no choice of coefficient recovers it. The same analytic closure that succeeds in the linear regime therefore fails here, whereas the single online-trained FNO reproduces both regimes within one model, albeit after re-training the FNO, however, in the following section we will present a single FNO model that can reproduce both the linear and nonlinear cases without re-training.

\begin{table}
  \centering
  \begin{tabular}{lcccc}
    \toprule
     & \multicolumn{2}{c}{Linear ($A = 10^{-3}$)}
     & \multicolumn{2}{c}{Nonlinear ($A = 10^{-1}$)} \\
    \cmidrule(lr){2-3} \cmidrule(lr){4-5}
    Field & Train & Test & Train & Test \\
    \midrule
    $E$        & $6.81\times10^{-3}$ & $8.78\times10^{-2}$
               & $5.30\times10^{-3}$ & $1.42\times10^{-1}$ \\
    $n$        & $1.76\times10^{-6}$ & $1.96\times10^{-6}$
               & $1.44\times10^{-4}$ & $2.86\times10^{-3}$ \\
    $u$        & $5.69\times10^{-3}$ & $1.04\times10^{-1}$
               & $8.64\times10^{-3}$ & $1.45\times10^{-1}$ \\
    $p$        & $8.33\times10^{-6}$ & $6.53\times10^{-6}$
               & $7.63\times10^{-4}$ & $1.25\times10^{-2}$ \\
    $\partial_x q$ & $6.05\times10^{-2}$ & $9.72\times10^{-2}$
               & $1.55\times10^{-1}$ & $1.01\phantom{\times10^{-0}}$ \\
    \bottomrule
  \end{tabular}
  \caption{%
    Time-averaged relative $L^2$ error of each evolved fluid moment between the FNO-closed fluid simulation and the kinetic ground truth, evaluated over the in-training (Train) and out-of-training (Test) windows. Window splits are $t_\mathrm{split}=75$ for the linear case and $t_\mathrm{split}=250$ for the nonlinear case.}
  \label{tab:single_sim_L2}
\end{table}

\begin{table}
  \centering
  \begin{tabular}{ccc}
    \toprule
    Threshold & Linear ($A=10^{-3}$) & Nonlinear ($A=10^{-1}$) \\
    \midrule
    $10\%$  & $1.40$ & $1.40$ \\
    $50\%$  & $45.1$ & $207$  \\
    $100\%$ & $55.4$ & $270$  \\
    \bottomrule
  \end{tabular}
  \caption{%
    Elapsed time $t_\mathrm{div}$ (in $\omega_{\mathrm{p}}^{-1}$), measured from the end of the training window, at which the time-averaged relative error in the electric field $E$ first exceeds the indicated threshold, for the linear and nonlinear single-simulation cases (training windows ending at $t=75$ and $t=250\,\omega_{\mathrm{p}}^{-1}$ respectively). The $10\%$ threshold is crossed almost immediately in both cases, but the linear case reaches $50\%$ and $100\%$ error far sooner than the nonlinear case, which does not reach $100\%$ within the simulated window.}
  \label{tab:single_sim_divtime}
\end{table}

\subsection{Generalising across Different Initial Amplitudes}
\label{sec:generalisation}

To test whether a single trained closure can capture both linear and nonlinear regimes within one model, we train an FNO on data drawn from a range of initial perturbation amplitudes spanning the linear-to-nonlinear transition. The training amplitudes are sampled with four points per decade in $\log A$, giving
\begin{equation}
    A_\mathrm{train} = 10^{(j/4 - 3)}, \qquad j \in \{0, 1, 2, \ldots, 8\}
\end{equation}
ranging from $A = 10^{-3}$ (strongly linear) to $A = 10^{-1}$ (strongly nonlinear). The choice of four points per decade is intended to provide reasonable coverage of the linear, transition, and nonlinear regimes, with sufficient density in the transition region where the qualitative dynamics change rapidly with amplitude.

We test the trained FNO on a denser grid of amplitudes, spread evenly throughout the 2 decades we include in the training set of simulations. Our test set  is therefore comprised of simulations with the initial perturbation amplitudes: $(7 \times 10^{-4}, 8 \times 10^{-4}, 9 \times 10^{-4}, 1 \times 10^{-3}, 2 \times 10^{-3}, \dots, 9 \times 10^{-2}, 1 \times 10^{-1}, 2 \times 10^{-1}, 3 \times 10^{-1})$. This test set includes amplitudes that coincide with training values, permitting direct in-sample evaluation, and amplitudes that do not, permitting evaluation of the closure's ability to interpolate between the regimes represented in the training data. Training data is mixed across amplitudes during training: at each iteration, snapshots are sampled uniformly across the training set, ensuring that no single amplitude dominates the loss signal.

\begin{figure}
    \centering
    \includegraphics[width=0.9\linewidth]{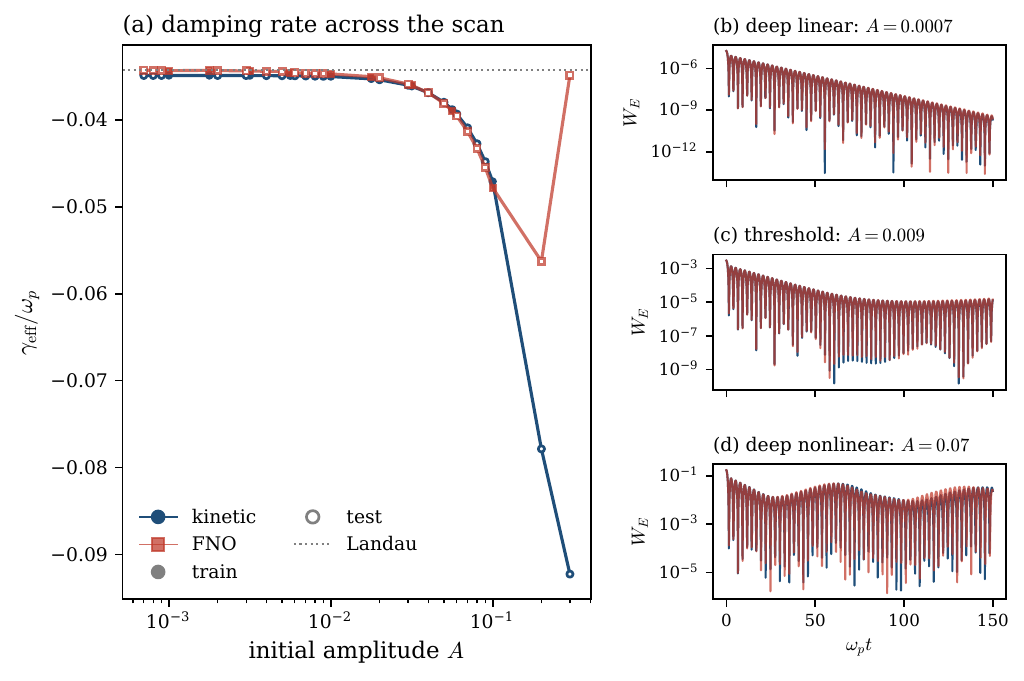}
    \caption{(a) Damping rate of the initial linear phase of each amplitude run, comparing the kinetic and fluid+FNO. Damping rate is calculated by a linear fit to the peaks of the $W_E$ plots over 0-15 $\omega_{\mathrm{p}} t$ of each simulation run. (b), (c), (d) are plots of $W_E$ over simulation time for a linear, threshold, and nonlinear case of Landau damping respectively.}
    \label{fig:general_summary_panel}
\end{figure}

\begin{figure}
    \centering
    \includegraphics[width=1.0\linewidth]{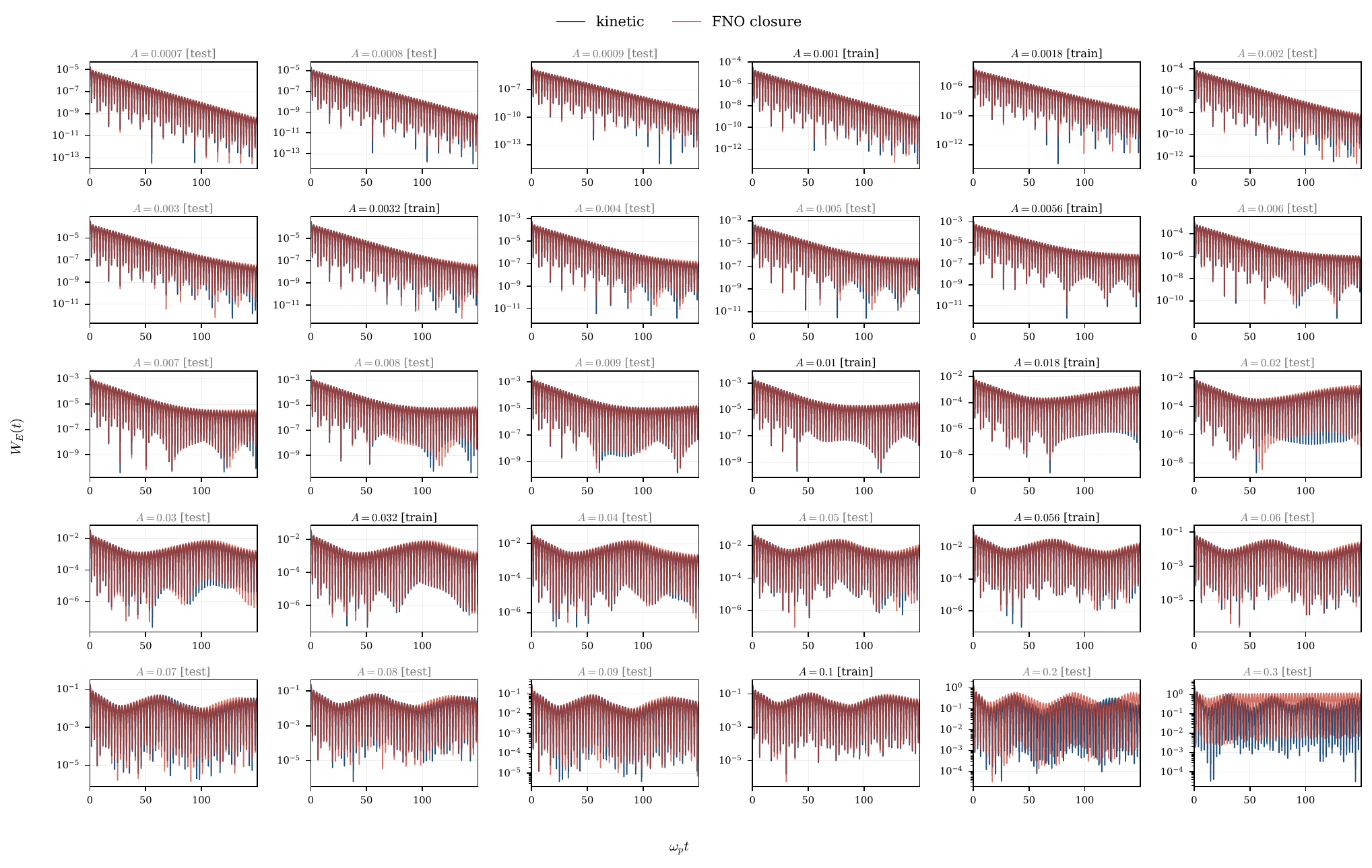}
    \caption{Electric-field energy $W_E(t)$ for every amplitude in the test grid, comparing the kinetic simulation (blue) with the online-tested generalised FNO closure (red); each panel is labelled by its initial amplitude and by whether that amplitude was included in training. All FNO curves are produced by online testing of the single trained closure from the same initial conditions as the corresponding kinetic run. Agreement is close across the linear, threshold and nonlinear regimes and at both trained and interpolated amplitudes; the two largest amplitudes ($A=0.2,\,0.3$), which lie above the training range, show the FNO envelope over-predicting the kinetic one, consistent with the extrapolation failure seen in Figure~\ref{fig:general_summary_panel}(a).}
    \label{fig:general_W_E}
\end{figure}

Figure~\ref{fig:general_summary_panel} summarises the performance of the trained closure across all test amplitudes. Panel~(a) shows the linear damping rate measured in the initial damping window of each run for both the kinetic simulation and the fluid+FNO simulation; the agreement is good across the majority of the amplitude range. The closure tracks the kinetic rate not only through the linear plateau but through the nonlinear enhancement of the damping as the amplitude increases, and interpolates accurately to test amplitudes that lie between training values. The agreement breaks down only for the two largest test amplitudes ($A=0.2$ and $0.3$), which lie beyond the maximum amplitude included in training: here the FNO rate departs sharply from the kinetic one, whereas test amplitudes that fall below the minimum training amplitude remain well captured. This asymmetry, a failure to extrapolate above the training envelope but reliable interpolation and extrapolation below it, demonstrates the need to span all dynamically relevant regimes when assembling the training set. Panels~(b)-(d) show the time evolution of $W_E$ for representative deep-linear, threshold and deep-nonlinear cases, confirming that a single trained model captures the qualitative evolution in all three regimes. The full set of per-amplitude $W_E(t)$ traces is collected in Figure~\ref{fig:general_W_E}, which shows close kinetic-FNO agreement at every trained and interpolated amplitude and the same over-prediction of the envelope at the two extrapolated amplitudes.

\begin{figure}
    \centering
    \includegraphics[width=0.8\linewidth]{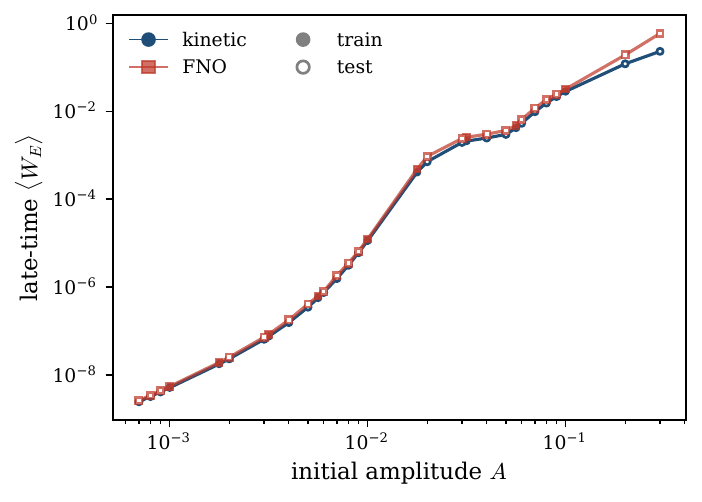}
    \caption{Saturation electric-field energy as a function of initial amplitude, comparing the kinetic and fluid+FNO simulations. The saturation value is computed as the time average of $W_E$ over the final $25\%$ of each simulation run, at which point the nonlinear cases have reached an approximate steady state. The fluid+FNO closure reproduces the saturated energy across the nonlinear regime; the linear cases, for which no true saturation is reached within the simulation window and the metric is correspondingly ill-defined, are nonetheless captured well.}
    \label{fig:general_saturation}
\end{figure}

Figure~\ref{fig:general_saturation} reports the saturation electric field energy for each amplitude run, computed as the time average of $W_E$ over the final 25\% of timesteps of each simulation. The fluid-FNO simulations reproduce the saturation behaviour observed in the kinetic simulations across the nonlinear regime; the linear cases, in which no true saturation is reached within the simulation window are still well captured by the FNO closure, despite the ill-defined nature of the saturation electric field energy for these cases.

\begin{figure}
    \centering
    \includegraphics[width=0.8\linewidth]{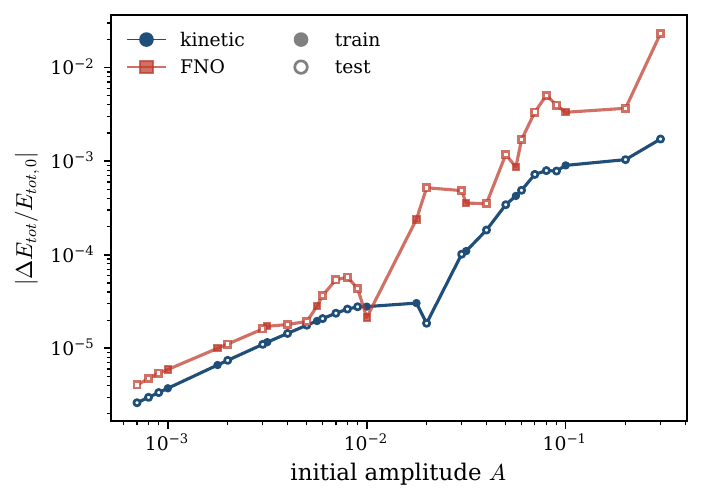}
    \caption{Increase of the total energy over time for the kinetic and fluid+FNO simulations across the amplitude scan, quantifying spurious energy injection or dissipation introduced by the closure. In the linear regime the fluid+FNO drift is comparable to the kinetic reference; in the deeply nonlinear regime it remains bounded, though roughly one to two orders of magnitude larger than the kinetic drift. Energy conservation is not imposed on the closure, so the bounded drift reflects consistency of the learned closure with the underlying equations rather than an enforced constraint.}
    \label{fig:general_conservation}
\end{figure}

Figure~\ref{fig:general_conservation} shows the increase in total energy, $E_{\mathrm{tot}}$, of both kinetic and fluid-FNO simulations over time. The total energy is a Vlasov invariant, and so will be conserved in the Vlasov solver kinetic simulations, however, when calculating the total energy from the truncated moment hierarchy, an increase is expected due to the loss of the higher-order moments. The fluid-FNO simulations exhibit comparable conservation behaviour to the kinetic reference in the linear regimes, while remaining bounded in the deeply nonlinear regimes, despite growing to roughly 1-2 order of magnitude larger than the kinetic results; indicating that the learned closure does not introduce significant spurious energy injection or dissipation over the simulation window. We note that the FNO closure is not constructed to enforce energy conservation explicitly, the level of conservation observed therefore reflects the consistency of the learned closure with the physical structure of the underlying equations rather than an imposed constraint.

\begin{figure}
    \centering
    \includegraphics[width=0.85\linewidth]{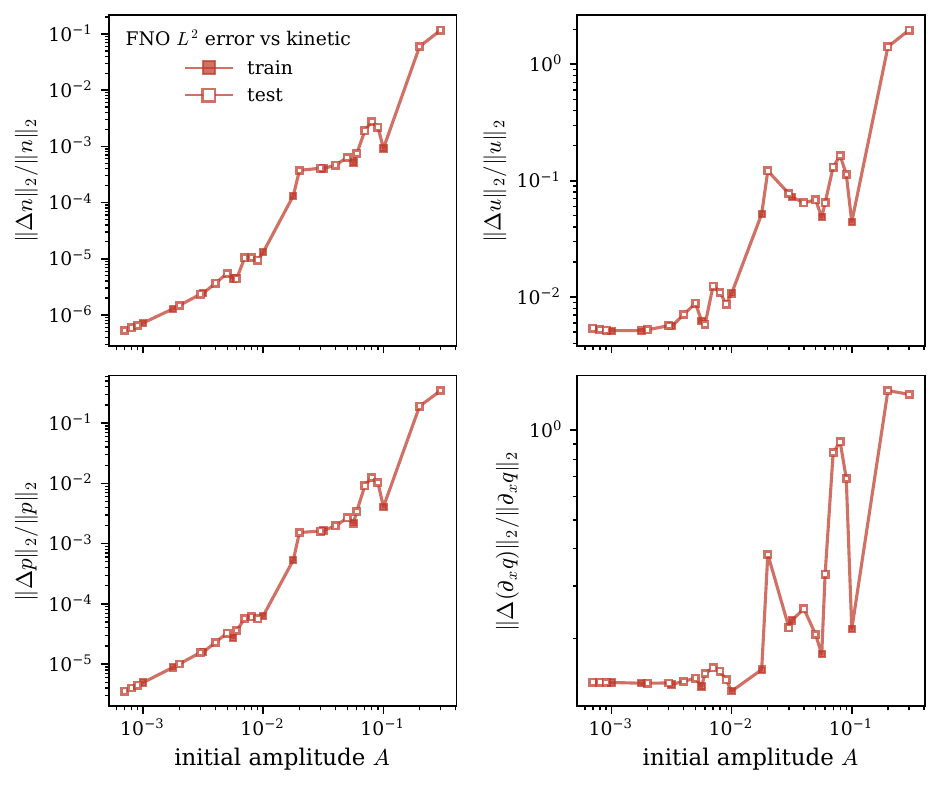}
    \caption{Relative $L^2$ error of each evolved moment's trajectory between the kinetic and fluid+FNO simulations, as a function of initial perturbation amplitude, for the single generalised FNO closure. One panel is shown per field ($E$, $n$, $u$, $p$, $\partial_x q$); filled markers denote amplitudes included in the training set and open markers the held-out test amplitudes. Errors increase monotonically with amplitude as the dynamics become more nonlinear, with training and test markers coinciding across the scan. The heat-flux-gradient error (bottom centre) reaches order unity as soon as the trapping regime is entered, whereas the resolved-moment errors rise more gradually.}
    \label{fig:general_l2}
\end{figure}

Figure~\ref{fig:general_l2} resolves the per-moment relative $L^2$ error against the kinetic reference as a function of initial amplitude, separated into training and test amplitudes. For every field the error rises smoothly with amplitude as the dynamics become more nonlinear, and the training and test markers lie essentially on top of one another across the whole scan, confirming that the closure interpolates between trained amplitudes without a generalisation gap. The errors in $E$, $n$, $u$ and $p$ climb from below $10^{-2}$ in the linear regime to order unity only at the largest amplitudes, with a characteristic step near $A\approx2\times10^{-2}$ where the system crosses into the trapping regime. The heat-flux-gradient error is the exception: it jumps to order unity as soon as the dynamics become nonlinear and remains there, mirroring the single-amplitude behaviour of Section~\ref{sec:nonlinear} and anticipating the effective-heat-flux discussion of Section~\ref{sec:heatflux}.

\subsection{The learned closure as an effective heat flux}
\label{sec:heatflux}

Because the online closure is never trained against the kinetic heat flux directly (Section~\ref{sec:online_method}), there is no guarantee, nor any requirement, that the heat-flux gradient it produces coincides with the one measured from the kinetic simulation. The closure is instead free to converge to whichever $\partial_x q$ best reproduces the evolution of the resolved moments under the specific fluid discretisation in use. We therefore distinguish the \emph{effective} heat flux learned by the online closure from the \emph{kinetic} heat flux computed as a velocity moment of the distribution function.

In the linear regime the two remain close. Table~\ref{tab:single_sim_L2} shows that the relative $L^2$ error of $\partial_x q$ against the kinetic value stays at the $10^{-1}$ level throughout the linear run, of the same order as the velocity and electric-field errors. In the nonlinear regime, by contrast, the effective and kinetic heat fluxes diverge markedly: the $\partial_x q$ error grows to order unity in the test window, an order of magnitude or more above the density and pressure errors, which remain at the $10^{-2}$ level (Table~\ref{tab:single_sim_L2}), while the electric-field energy, including its saturation and bounce oscillations, is well reproduced (Section~\ref{sec:nonlinear}). The spatial structure of the discrepancy is visible in the $\partial_x q$ row of the contour plots (Figure~\ref{fig:nonlinear_contour}). The closure thus reproduces the observable fluid dynamics while producing a heat flux that departs substantially from the kinetic one.

This behaviour is consistent with the truncation of the moment hierarchy, an interpretation we introduced in Section~\ref{sec:online_method}. The kinetic system carries the full velocity dependence of the distribution function, whereas the fluid model retains only moments up to the pressure; the influence of the higher moments, and of the fine velocity-space structure generated by phase mixing, has been projected out. The Mori-Zwanzig formalism shows that, in general, such projected influence reappears as a history-dependent term acting on the resolved variables \citep{gouasmi_priori_2017, parish_nonmarkovian_2017}, and this provides a natural interpretation of what the online closure has learned: not the bare kinetic heat flux, but an effective flux that compensates for the missing higher-moment information so as to reproduce the evolution of the resolved fields. We emphasise, however, that the specific effective flux obtained is not a unique or scheme-independent object. In this language the Hammett-Perkins closure (\ref{eq:hp}) is the Markovian, linear limit of the effective flux: a memoryless, spatially non-local operator that captures the linear response but carries none of the history the trapping regime requires. The online closure can be read as a nonlinear, non-Markovian generalisation of the same object, its spectral layers acting as a learned counterpart of the $\mathrm{i}k/|k|$ multiplier, consistent with the growth of its memory reliance into the nonlinear regime (Section~\ref{sec:interpretability}). Since it is defined only through its effect on the discretised resolved-moment dynamics, it is expected to depend on the fluid discretisation, the time-integration scheme, and the training distribution, and a different solver or training set would in general yield a quantitatively different effective closure. We therefore intend this as a qualitative interpretation rather than a recovery of the exact Mori-Zwanzig memory term. It is consistent with the memory-window design of the closure (Section~\ref{sec:fluid_model}) and with the sensitivity analysis of Section~\ref{sec:interpretability}, which shows the closure's dependence on the $\partial_x q$ history growing, and decoupling from the lower moments, as the dynamics become nonlinear.

The contrast with the offline closure is instructive. Because the offline model is trained to regress $\partial_x q_{\mathrm{kin}}$ directly (Section~\ref{sec:offline}), it reproduces the kinetic heat flux more closely by construction; yet, in our matched comparison, this closer fidelity to the kinetic flux did not translate into better reproduction of the resolved-moment dynamics under online deployment (Sections~\ref{sec:linear} and~\ref{sec:nonlinear}). We do not read this as evidence that offline training is inherently inferior, offline closures can perform well under other configurations, including for this same problem \citep{huang_machine-learning_2025}. Rather, it suggests that, at least under identical hyperparameters and training data, reproducing the kinetic heat flux pointwise is not by itself sufficient for an accurate closed fluid model; what matters is the closure's effect on the retained moments, which online training optimises directly.

\subsection{What the closure has learned}
\label{sec:interpretability}

Having established that the closure reproduces the kinetic dynamics and generalises across amplitude (Sections~\ref{sec:linear}-\ref{sec:generalisation}), we now ask how it does so. The FNO closure is a black-box model, and such a closure could in principle reproduce a trajectory for the wrong reason, by extrapolating its own recent heat-flux output forward in time rather than computing the heat flux from the state of the plasma; such a model would not be expected to generalise across amplitudes as ours does. Two questions follow, first, does the closure compute a genuine relation from the resolved moment fields ($n$, $u$, $p$, $E$) to the heat-flux gradient, or does it merely continue its own $\partial_x q$ history? Second, does the memory window introduced in Section~\ref{sec:fluid_model} carry information the closure actually uses?

We probe both by measuring how the predicted heat flux gradient of the FNO responds to its inputs. The primary diagnostic is an ablation test, where we replace a single input field, or a group of them, with its average value, which removes the spatial and temporal variation, and record the resulting change in $\partial_x q$. A large change signals a strong dependence on that input. We complement this with the Hilbert-Schmidt independence criterion (HSIC) \citep{gretton_measuring_2005}, a measure of statistical dependence that, unlike a correlation coefficient, captures nonlinear as well as linear relationships. It generalises covariance by computing it in a transformed (kernel) feature space, so that it detects association of any functional form and vanishes only when the two quantities are statistically independent. Both diagnostics act on the deployed closure over representative data; since the ablation test displaces the inputs from the states the closure normally sees, we read the results as qualitative trends rather than precise values.

\begin{figure}
  \centering
  \includegraphics[width=0.9\textwidth]{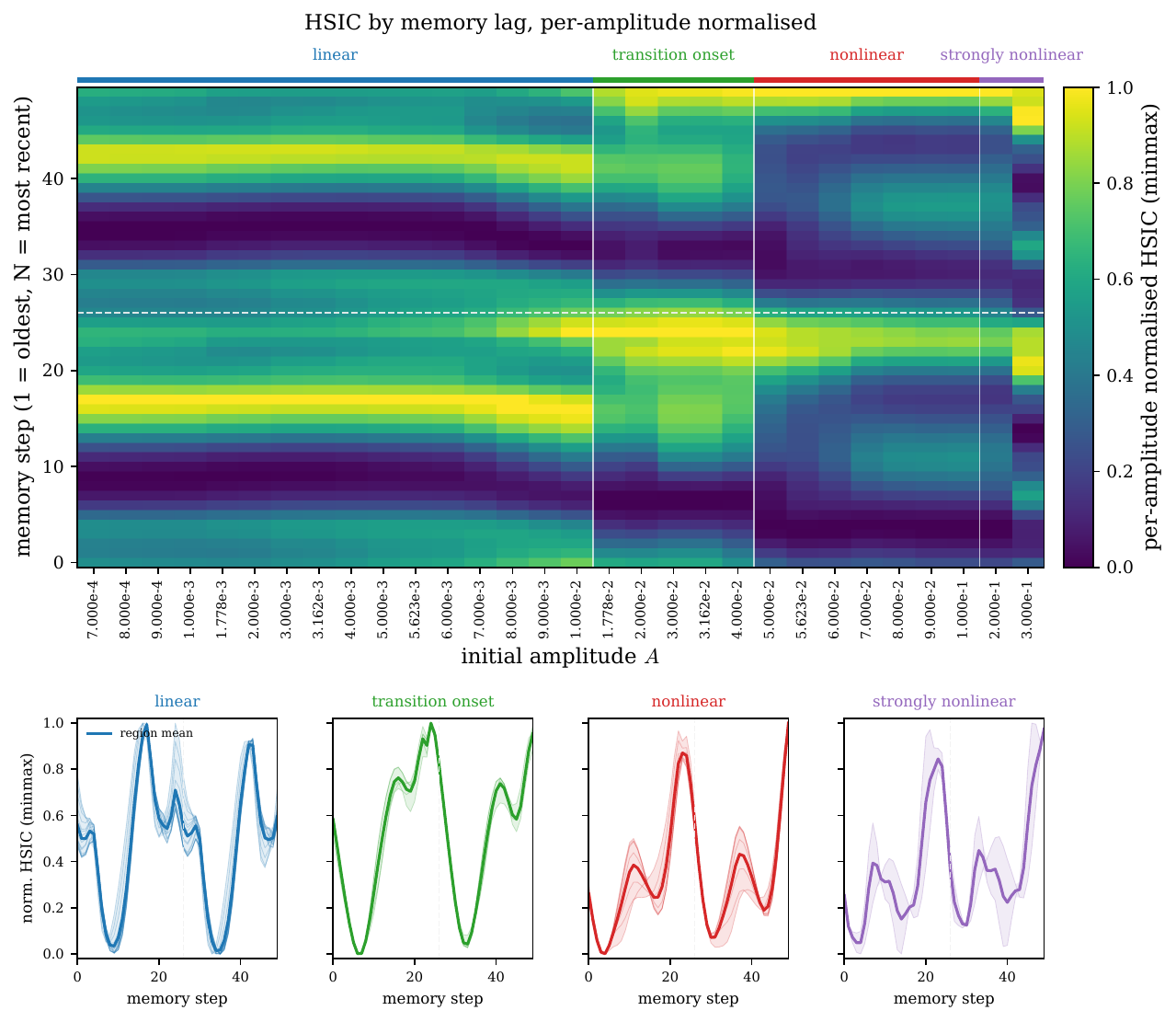}
  \caption{Strength of the closure's dependence on its memory inputs, as a function of how far back in time the input lies (vertical axis, the memory lag) and of initial perturbation amplitude (horizontal axis). Dependence is measured by the HSIC (a statistical association measure; see text) between each past field and the predicted heat-flux gradient, and is rescaled to $[0,1]$ separately at each amplitude so the lag pattern can be compared across amplitudes. The dependence concentrates at time lags spaced by the oscillation period of the electric-field energy (dashed line), showing that the closure reads the phase of the oscillation from its memory rather than an undifferentiated recent history. The pattern also changes character across the linear, threshold and nonlinear regimes (left to right).}
  \label{fig:hsic_lag}
\end{figure}

The memory window carries structured rather than uniform information (Figure~\ref{fig:hsic_lag}). The dependence on the past fields concentrates at particular lags, spaced by the oscillation period of the electric field energy, so the closure reads the phase of the underlying oscillation from its memory rather than an undifferentiated recent history; the lag structure additionally differs between the linear, threshold and nonlinear regimes. This structure should be read with a caveat intrinsic to the measure. Because the resolved fields oscillate at close to the plasma frequency, each past field is autocorrelated with the present state, which is in turn related to the target, so a given lag may register a high dependence either because the closure draws on it or because the field there is a proxy for the present state. The measured pattern is thus a combination of the true memory-dependence and the autocorrelation of the fields, which HSIC alone cannot separate; the autocorrelation of an oscillation peaks at lags spaced by its period, the same spacing we attribute to the closure reading the oscillation phase, so here the two reinforce rather than distinguish one another. What the autocorrelation cannot produce is the response to the ablation test below, in which the physical state is removed while the heat-flux history is retained.

\begin{figure}
  \centering
  \includegraphics[width=0.8\textwidth]{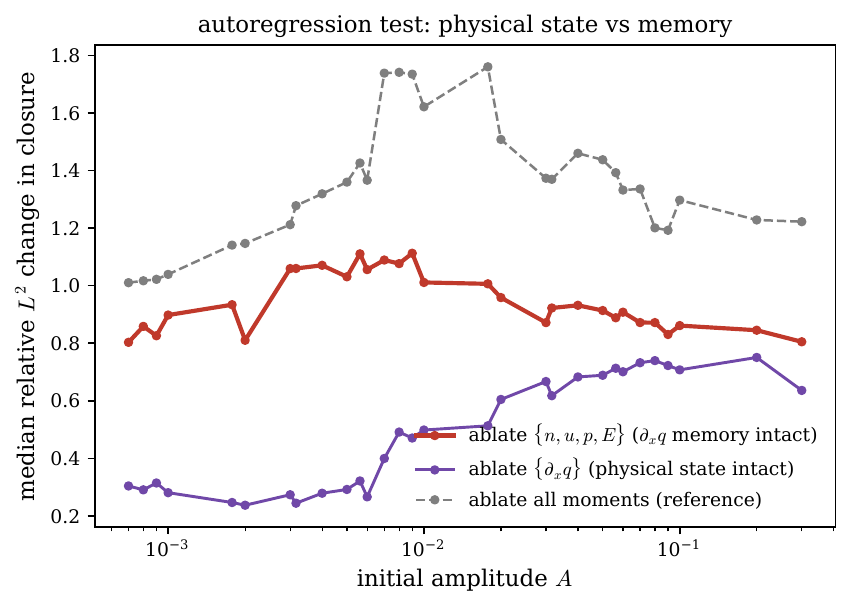}
  \caption{Test of whether the closure reads the plasma state or merely continues its own output. Each curve shows how much the predicted heat-flux gradient changes (median relative $L^2$ change) when a set of inputs is replaced by its average value, as a function of initial amplitude. Replacing the physical moment fields $\{n,u,p,E\}$ while keeping the heat-flux history intact (solid) changes the prediction strongly at every amplitude, and by more than removing the heat-flux history alone: the closure depends on the plasma state and is not simply extrapolating its past output. The overall dependence on the moment inputs (dashed) grows as the dynamics become nonlinear.}
  \label{fig:autoreg}
\end{figure}

The ablation test shows the closure to compute a genuine relation from the plasma state to the heat flux, rather than continuing its own output (Figure~\ref{fig:autoreg}). Replacing the physical moment fields $\{n,u,p,E\}$ with their averages while leaving the $\partial_x q$ history intact changes the prediction substantially at every amplitude, and by more than removing the $\partial_x q$ history alone; a closure extrapolating its past output would be almost unaffected by the loss of the physical state. This dependence on the moment fields strengthens rather than weakens as the dynamics become nonlinear, and taken with the generalisation of Section~\ref{sec:generalisation} indicates a transferable physical relation rather than memorised trajectories.

\begin{figure}
  \centering
  \begin{subfigure}{0.49\textwidth}
    \includegraphics[width=\textwidth]{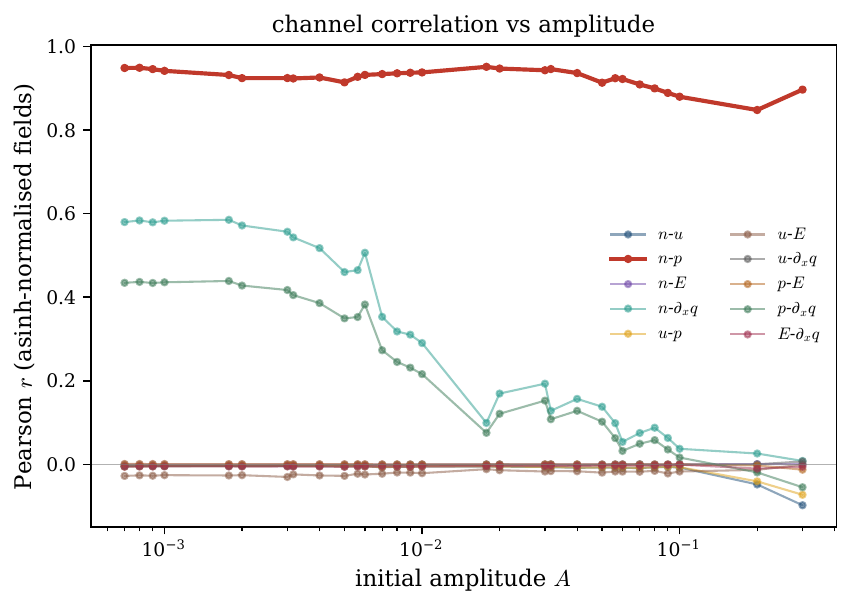}
    \caption{}
  \end{subfigure}
  \hfill
  \begin{subfigure}{0.49\textwidth}
    \includegraphics[width=\textwidth]{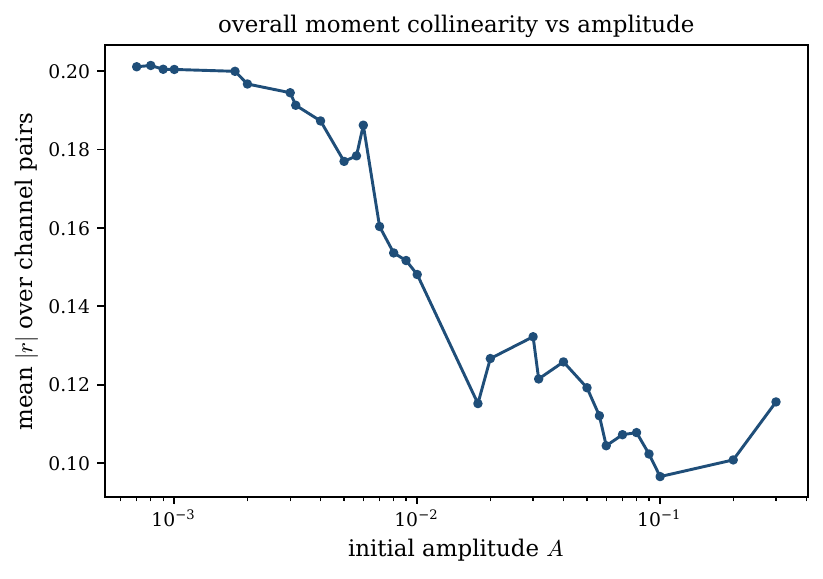}
    \caption{}
  \end{subfigure}
  \caption{How closely the input fields track one another, as a function of initial amplitude (computed on the fields as supplied to the network). (a) Correlation between pairs of input fields: values near $\pm1$ mean the two carry nearly the same information, values near $0$ mean they are unrelated. The density-pressure pair (emphasised) stays near $0.9$ at all amplitudes, the two are nearly interchangeable, while the heat-flux history becomes progressively unrelated to $n$ and $p$ as the dynamics become nonlinear. (b) The average correlation across all pairs, summarising how much the inputs overlap overall; its decrease with amplitude is driven by the heat-flux history decoupling from the lower moments, not by any change in the density-pressure pair.}
  \label{fig:correlation}
\end{figure}

\begin{figure}
  \centering
  \includegraphics[width=0.8\textwidth]{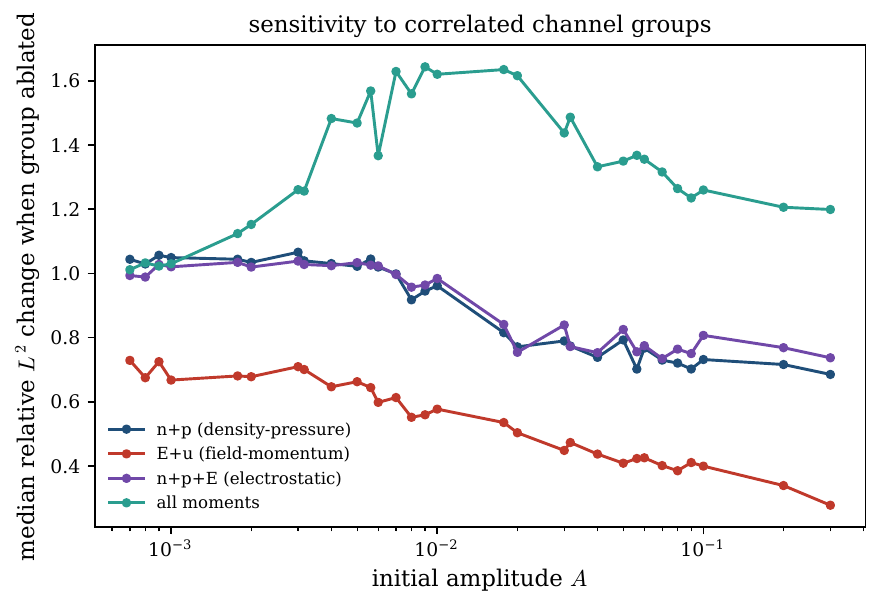}
  \caption{Dependence of the closure on groups of related inputs, measured as the median relative $L^2$ change in the predicted heat-flux gradient when each group is replaced by its average, versus initial amplitude. The $\{n,p\}$ and $\{n,p,E\}$ curves almost coincide, so adding the electric field to the density-pressure group makes little difference: $E$ carries essentially no information the density does not (as expected from Poisson's equation). The curve for all moment fields together rises with amplitude, showing that the closure leans more heavily on the resolved state, not less, as the dynamics become nonlinear.}
  \label{fig:group}
\end{figure}

\begin{figure}
  \centering
  \includegraphics[width=0.8\textwidth]{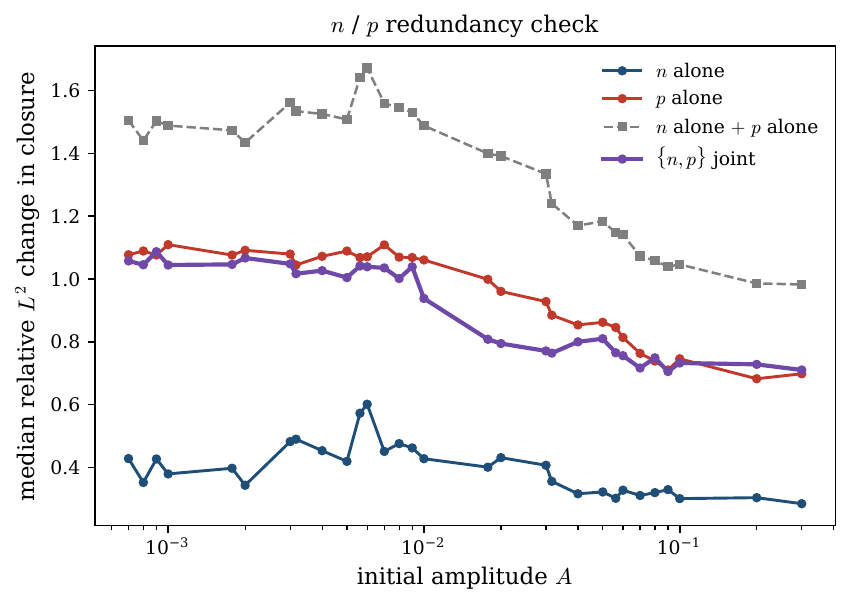}
  \caption{Which of density and pressure the closure actually uses. Curves show the change in the predicted heat-flux gradient when density alone, pressure alone, or both together are replaced by their averages, versus initial amplitude, together with the sum of the two single-field values. Removing both together has an effect close to removing pressure alone, and smaller than the two single-field effects added together: pressure carries most of the shared density-pressure information, and density adds little once pressure is present.}
  \label{fig:superadd}
\end{figure}

Among the resolved fields the closure draws primarily on the pressure. Density and pressure are nearly proportional across the amplitude scan (Figure~\ref{fig:correlation}a, correlation $\approx 0.9$) and so carry almost the same information, and the ablation tests identify pressure as the field that supplies it (Figure~\ref{fig:superadd}): removing density and pressure together changes the prediction by little more than removing pressure alone, leaving density largely redundant once pressure is present. The electric field is redundant with the density in the same way, adding little to the $\{n,p\}$ group when included (Figure~\ref{fig:group}), as the electrostatic constraint requires: Poisson's equation~(\ref{eq:poisson}) ties $E$ to the density through $\partial_x E = \frac{q_{\mathrm{e}}}{\epsilon_0}(n_{0}-n)$, so $E$ carries little that the density does not.

Finally, the diagnostics indicate why the memory is needed, beyond confirming that it is used. As the amplitude increases into the nonlinear regime the heat-flux history becomes at once more important to the closure (Figure~\ref{fig:group}) and less related to the instantaneous lower moments, its correlation with $n$ and $p$ falling towards zero (Figure~\ref{fig:correlation}). The heat flux thus acquires dynamics that cannot be reconstructed from the present values of the other moments, the signature of phase mixing, precisely the regime in which a memory-dependent, non-Markovian closure is necessary. The measured behaviour of the closure therefore mirrors the physical reasoning that motivated its finite-memory design (Section~\ref{sec:fluid_model}). This growing reliance on the heat-flux history is the same trend invoked in Section~\ref{sec:heatflux}: the closure's departure from the kinetic heat flux in the nonlinear regime, and its increasing dependence on memory there, are two readings of one fact, that the effective flux becomes a genuinely history-dependent quantity as the truncated higher-moment dynamics grow in importance.

\section{Discussion}
\label{sec:discussion}

The results of the previous section establish three claims. First, training a Fourier Neural Operator closure online, within a differentiable fluid solver, produces a closure that is numerically stable when subsequently deployed in independent fluid simulations and that reproduces the kinetic dynamics of one-dimensional electrostatic Landau damping in both linear and nonlinear regimes. Second, a single online-trained FNO can generalise across initial perturbation amplitudes that lie between those represented in the training set. Third, in the nonlinear regime the heat flux learned by the online-trained closure does not exactly reproduce the heat flux measured directly from the kinetic simulation.

Taken together, these results establish online training as a viable route to producing learned closures that operate stably and accurately within fluid simulations and that generalise across the range of initial conditions one would encounter in a typical large-scale modelling application. To our knowledge, this work provides the first systematic demonstration that a learned plasma closure can generalise across a span of initial conditions while remaining stable in online deployment.

Several limitations of the present study suggest directions for further work. The online training procedure is substantially more memory-intensive than offline training, owing to the need to backpropagate through the fluid rollout. This high-memory requirement places constraints on the rollout length and batch size achievable on standard hardware and motivates an investigation of memory-efficient automatic differentiation strategies such as gradient checkpointing, or implicit differentiation. The choice of the FNO architecture, while well-motivated for the periodic, low-wavenumber dynamics of Landau damping, has not been compared against alternative architectures trained under the same online procedure; such a comparison would clarify the extent to which our results reflect properties of the FNO specifically as opposed to properties of online training more generally. The geometry of our problem, one-dimensional, electrostatic, with fixed wavenumber, is also deliberately simple. Extending the online training method to fully electromagnetic, multi-dimensional settings, and to phenomena such as magnetic reconnection or plasma turbulence \citep{2019JPlPh..85e9006P}, will require addressing both the increased dimensionality of the inputs and outputs and the much greater computational cost of the underlying kinetic simulations.

The spectral truncation of the FNO, which retains only the lowest Fourier modes in each layer, is well matched to the single-mode, low-wavenumber dynamics considered here, but may become a limitation in richer settings where several wavemodes are dynamically active or where important structure resides at higher wavenumbers. The number of retained modes can of course be increased, but at the cost of additional parameters and training expense, and the appropriate mode content is ultimately problem-dependent. Extending the method to multi-mode or broadband problems will therefore require matching the retained spectrum to the range of scales that carry the physics of interest, and assessing how the closure behaves for modes whose damping rates differ substantially across the retained band.

Despite these limitations, the present results are a concrete step towards machine-learned plasma closures suitable for large-scale fluid simulation. Where kinetic simulations are too expensive to run repeatedly but the relevant physical regimes are spanned by a tractable subset of fully-kinetic runs, an online-trained closure derived from that subset offers a route to fluid-cost simulations that retain the kinetic effects most relevant to the regime of interest.

\section{Acknowledgements}

This research utilised Queen Mary's Apocrita HPC facility, supported by QMUL Research-IT. (DOI: \url{http://doi.org/10.5281/zenodo.438045})
This research was supported by a Queen Mary University of London Principal's Award Studentship.
OP acknowledges support by the FIS2 Starting Grant FIS-2023-00246 “PhAse-sPAce cOmplexity in turbulent nearly-reversible plasmas (PAPAO)” (CUP B53C24009610001) funded by the Italian Ministry of University and Research. 
\bibliographystyle{jpp}
\bibliography{references}

\end{document}